\documentclass[12pt]{article}
\usepackage{multirow}
\usepackage{latexsym} 
\usepackage{color}
\usepackage{epsfig}
\usepackage{amsmath}
\usepackage{color}
\usepackage{colortbl}
\usepackage{hhline}
\usepackage{amssymb}
\usepackage{times}
\usepackage{cite}
\usepackage{array}

\definecolor{rltred}{rgb}{0.75,0,0}
\definecolor{rltgreen}{rgb}{0,0.5,0}
\definecolor{rltblue}{rgb}{0,0,0.75}

\newlength{\dinwidth}
\newlength{\dinmargin}
\begin{document}

\newcommand{\pom}{{I\!\!P}}
\newcommand{\reg}{{I\!\!R}}
\newcommand{\slowpi}{\pi_{\mathit{slow}}}
\newcommand{\fiidiii}{F_2^{D(3)}}
\newcommand{\fiidiiiarg}{\fiidiii\,(\beta,\,Q^2,\,x)}
\newcommand{\n}{1.19\pm 0.06 (stat.) \pm0.07 (syst.)}
\newcommand{\nz}{1.30\pm 0.08 (stat.)^{+0.08}_{-0.14} (syst.)}
\newcommand{\fiidiiiful}{F_2^{D(4)}\,(\beta,\,Q^2,\,x,\,t)}
\newcommand{\fiipom}{\tilde F_2^D}
\newcommand{\ALPHA}{1.10\pm0.03 (stat.) \pm0.04 (syst.)}
\newcommand{\ALPHAZ}{1.15\pm0.04 (stat.)^{+0.04}_{-0.07} (syst.)}
\newcommand{\fiipomarg}{\fiipom\,(\beta,\,Q^2)}
\newcommand{\pomflux}{f_{\pom / p}}
\newcommand{\nxpom}{1.19\pm 0.06 (stat.) \pm0.07 (syst.)}
\newcommand {\gapprox}
   {\raisebox{-0.7ex}{$\stackrel {\textstyle>}{\sim}$}}
\newcommand {\lapprox}
   {\raisebox{-0.7ex}{$\stackrel {\textstyle<}{\sim}$}}
\def\gsim{\,\lower.25ex\hbox{$\scriptstyle\sim$}\kern-1.30ex%
\raise 0.55ex\hbox{$\scriptstyle >$}\,}
\def\lsim{\,\lower.25ex\hbox{$\scriptstyle\sim$}\kern-1.30ex%
\raise 0.55ex\hbox{$\scriptstyle <$}\,}
\newcommand{\pomfluxarg}{f_{\pom / p}\,(x_\pom)}
\newcommand{\dsf}{\mbox{$F_2^{D(3)}$}}
\newcommand{\dsfva}{\mbox{$F_2^{D(3)}(\beta,Q^2,x_{I\!\!P})$}}
\newcommand{\dsfvb}{\mbox{$F_2^{D(3)}(\beta,Q^2,x)$}}
\newcommand{\dsfpom}{$F_2^{I\!\!P}$}
\newcommand{\gap}{\stackrel{>}{\sim}}
\newcommand{\lap}{\stackrel{<}{\sim}}
\newcommand{\fem}{$F_2^{em}$}
\newcommand{\tsnmp}{$\tilde{\sigma}_{NC}(e^{\mp})$}
\newcommand{\tsnm}{$\tilde{\sigma}_{NC}(e^-)$}
\newcommand{\tsnp}{$\tilde{\sigma}_{NC}(e^+)$}
\newcommand{\st}{$\star$}
\newcommand{\sst}{$\star \star$}
\newcommand{\ssst}{$\star \star \star$}
\newcommand{\sssst}{$\star \star \star \star$}
\newcommand{\tw}{\theta_W}
\newcommand{\sw}{\sin{\theta_W}}
\newcommand{\cw}{\cos{\theta_W}}
\newcommand{\sww}{\sin^2{\theta_W}}
\newcommand{\cww}{\cos^2{\theta_W}}
\newcommand{\trm}{m_{\perp}}
\newcommand{\trp}{p_{\perp}}
\newcommand{\trmm}{m_{\perp}^2}
\newcommand{\trpp}{p_{\perp}^2}
\newcommand{\alp}{\alpha_s}

\newcommand{\alps}{\alpha_s}
\newcommand{\sqrts}{$\sqrt{s}$}
\newcommand{\LO}{$O(\alpha_s^0)$}
\newcommand{\Oa}{$O(\alpha_s)$}
\newcommand{\Oaa}{$O(\alpha_s^2)$}
\newcommand{\PT}{p_{\perp}}
\newcommand{\JPSI}{J/\psi}
\newcommand{\sh}{\hat{s}}
\newcommand{\uh}{\hat{u}}
\newcommand{\MP}{m_{J/\psi}}
\newcommand{\PO}{I\!\!P}
\newcommand{\xbj}{x}
\newcommand{\xpom}{x_{\PO}}
\newcommand{\ttbs}{\char'134}
\newcommand{\xpomlo}{3\times10^{-4}}  
\newcommand{\xpomup}{0.05}  
\newcommand{\dgr}{^\circ}
\newcommand{\pbarnt}{\,\mbox{{\rm pb$^{-1}$}}}
\newcommand{\QQ}{$QQ~$\/\,}
\newcommand{\LQ}{$LQ~$\/\,}
\newcommand{\LL}{$LL~$\/\,}
\newcommand{\mev}{\,\mbox{MeV}}
\newcommand{\WBoson}{\mbox{$W$}}
\newcommand{\fbarn}{\,\mbox{{\rm fb}}}
\newcommand{\fbarnt}{\,\mbox{{\rm fb$^{-1}$}}}
%
%
\newcommand{\mc}{MC }
\newcommand{\mcs}{MCs }
\newcommand{\qsq}{\ensuremath{Q^2} }
\newcommand{\gev}{\ensuremath{\,\mathrm{GeV}} }
\newcommand{\gevsq}{\ensuremath{\,\mathrm{GeV}^2} }
\newcommand{\et}{\ensuremath{E_t^*} }
\newcommand{\rap}{\ensuremath{\eta^*} }
\newcommand{\gp}{\ensuremath{\gamma^*}p }
\newcommand{\dsiget}{\ensuremath{{\rm d}\sigma_{ep}/{\rm d}E_t^*} }
\newcommand{\dsigrap}{\ensuremath{{\rm d}\sigma_{ep}/{\rm d}\eta^*} }
\def\Journal#1#2#3#4{{#1} {\bf #2} (#3) #4}
\def\NCA{\em Nuovo Cimento}
\def\NIM{\em Nucl. Instrum. Methods}
\def\NIMA{{\em Nucl. Instrum. Methods} {\bf A}}
\def\NPB{{\em Nucl. Phys.}   {\bf B}}
\def\PLB{{\em Phys. Lett.}   {\bf B}}
\def\PRL{\em Phys. Rev. Lett.}
\def\PRD{{\em Phys. Rev.}    {\bf D}}
\def\ZPC{{\em Z. Phys.}      {\bf C}}
\def\EJC{{\em Eur. Phys. J.} {\bf C}}
\def\CPC{\em Comp. Phys. Commun.}
\def\jets{photon plus jet~}
\def\nojet{photon plus no-jets~}
\def\orjet{photon plus no-jets (jet)~}
\def\ORJET{Photon plus no-Jets (Jet)~}
\def\NOJET{Photon plus no-jets~}
\def\JETS{Photon plus jet~}
\def\SumMC{signal MC~}
\def\summc{signal MC}
\begin{titlepage}

\noindent
%
\begin{flushleft}
{\tt DESY 07-147    \hfill    ISSN 0418-9833} \\
{\tt October 2007}                  \\
\end{flushleft}

\vspace{2cm}

\begin{center}
\begin{Large}

{\bf Measurement of Isolated Photon Production in \\
Deep-Inelastic Scattering at HERA}

\vspace{2cm}

H1 Collaboration

\end{Large}
\end{center}

\vspace{2cm}

\begin{abstract}
\noindent
The production of isolated photons in deep-inelastic scattering
$ep\rightarrow e \gamma X$ is measured with the H1 detector at HERA.
The measurement is performed in the kinematic range of
 negative four-momentum transfer squared  $4<Q^2<150 $~GeV$^2$ and a 
 mass of the hadronic system  $W_X>50$~GeV. 
The analysis is based on a total integrated luminosity of $227$~pb$^{-1}$.
The production cross section of isolated
photons with a transverse energy in the range  $3 < E_T^\gamma < 10$~GeV
 and  pseudorapidity range  $-1.2 < \eta^\gamma < 1.8$
 is measured as a function of  $E_T^\gamma$, $\eta^\gamma$ and $Q^2$.
Isolated photon cross sections are also measured for events with no jets or
at least one hadronic jet.
The measurements are compared with predictions from Monte Carlo generators modelling the photon radiation from the quark and
the electron lines, as well as with calculations at leading and next to leading order in the strong coupling. The
predictions significantly underestimate the measured cross sections.

\end{abstract}

\begin{center}
Submitted to Eur.  Phys. J. {\bf C}
\end{center}

\end{titlepage}

%
%
%
\begin{flushleft}

F.D.~Aaron$^{5,49}$,           
A.~Aktas$^{11}$,               
C.~Alexa$^{5}$,                
V.~Andreev$^{25}$,             
B.~Antunovic$^{26}$,           
S.~Aplin$^{11}$,               
A.~Asmone$^{33}$,              
A.~Astvatsatourov$^{4}$,       
S.~Backovic$^{30}$,            
A.~Baghdasaryan$^{38}$
% ...

\bigskip{\it
 $ ^{1}$ I. Physikalisches Institut der RWTH, Aachen, Germany$^{ a}$ \\
 $ ^{2}$ Vinca  Institute of Nuclear Sciences, Belgrade, Serbia \\
% ...
$ ^{4}$ Inter-University Institute for High Energies ULB-VUB, Brussels;
          Universiteit Antwerpen, Antwerpen; Belgium$^{ c}$ \\
 $ ^{5}$ National Institute for Physics and Nuclear Engineering (NIPNE) ,
          Bucharest, Romania \\
% ...
$ ^{11}$ DESY, Hamburg, Germany \\
% ...
$ ^{25}$ Lebedev Physical Institute, Moscow, Russia$^{ e}$ \\
 $ ^{26}$ Max-Planck-Institut f\"ur Physik, M\"unchen, Germany \\
% ...
$ ^{30}$ Faculty of Science, University of Montenegro,
          Podgorica, Montenegro$^{ e}$ \\
% ...
$ ^{33}$ Dipartimento di Fisica Universit\`a di Roma Tre
          and INFN Roma~3, Roma, Italy \\
% ...
$ ^{38}$ Yerevan Physics Institute, Yerevan, Armenia \\
% ...
$ ^{49}$ Also at Faculty of Physics, University of Bucharest,
          Bucharest, Romania \\
% ...
$ ^a$ Supported by the Bundesministerium f\"ur Bildung und Forschung, FRG,
      under contract numbers 05 H1 1GUA /1, 05 H1 1PAA /1, 05 H1 1PAB /9,
      05 H1 1PEA /6, 05 H1 1VHA /7 and 05 H1 1VHB /5 \\
% ...
$ ^c$ Supported by FNRS-FWO-Vlaanderen, IISN-IIKW and IWT
      and  by Interuniversity
Attraction Poles Programme,
      Belgian Science Policy \\
% ...
$ ^e$ Supported by the Deutsche Forschungsgemeinschaft \\
% ...
}
\end{flushleft}

\newpage
\section{Introduction}
\noindent
Isolated photons originating from the hard interaction in high energy 
collisions involving hadrons are a sensitive probe
of perturbative Quantum Chromodynamics (QCD)~\cite{fontannaz,joerg}, as the photons are largely insensitive to the effects of hadronisation.
A good understanding of the Standard Model (SM)  production mechanism of
 isolated photons is also important for 
searches of new particles decaying to photons at hadron colliders.

The production of isolated photons\footnote{Photons coupling to the interacting partons are  often called ``prompt'' in contrast to photons from hadron decays or those emitted by leptons.}  has been studied at various experiments. 
 Cross sections measured in fixed target $pN$ experiments (e.g. \cite{E706}) show  a steeper decrease with photon transverse momentum, 
$P_T^\gamma$, than     predicted by next-to-leading order (NLO) QCD calculations.
The CDF~\cite{CDF0} and D0~\cite{Abbott:1999kd} experiments at the Tevatron   have measured the isolated photon production cross section in $p\bar{p}$ 
collisions. 
Whereas D0 finds good agreement with a NLO QCD  calculation, the   CDF data
   show  a somewhat steeper $P_T^\gamma$ dependence than predicted.
 Measurements of the photon production in $e^+e^-$ collisions have also been
performed at LEP \cite{lepphotons}.  
At HERA, prompt
 photon cross
sections  have been measured by the H1 and ZEUS experiments~\cite{ZEUS,H1,Brownson} 
in photoproduction, where the negative four-momentum transfer
squared $Q^2$ of the exchanged virtual photon is close to zero,
and showed reasonable agreement with NLO calculations.
An analysis of the  isolated photon cross section in deep-inelastic scattering (DIS) with $Q^2$ larger than $35$~GeV$^2$ has  been published by ZEUS~\cite{ZEUSdis}. 
 
\begin{figure}[Hhh]
  \begin{center}
   \hspace{0.04\textwidth}
    \includegraphics[width=0.30\textwidth]{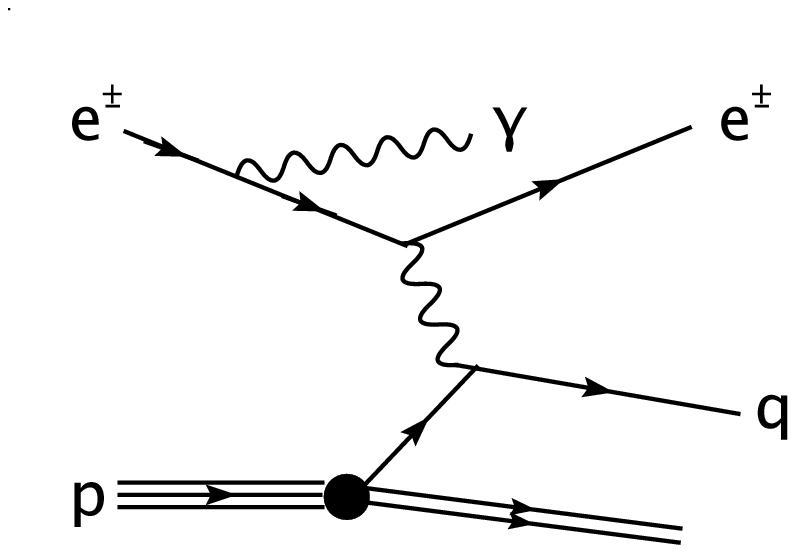}
    \hspace{0.04\textwidth}
    \includegraphics[width=0.30\textwidth]{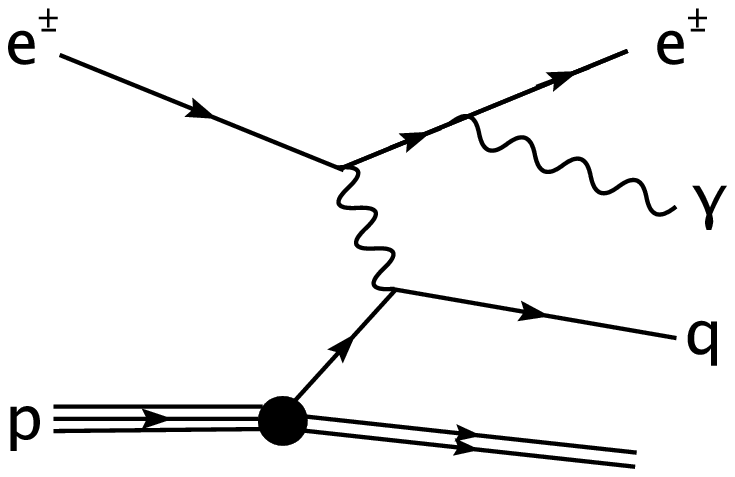}
  \end{center}
  \begin{center}
    \hspace{0.04\textwidth}
    \includegraphics[width=0.30\textwidth]{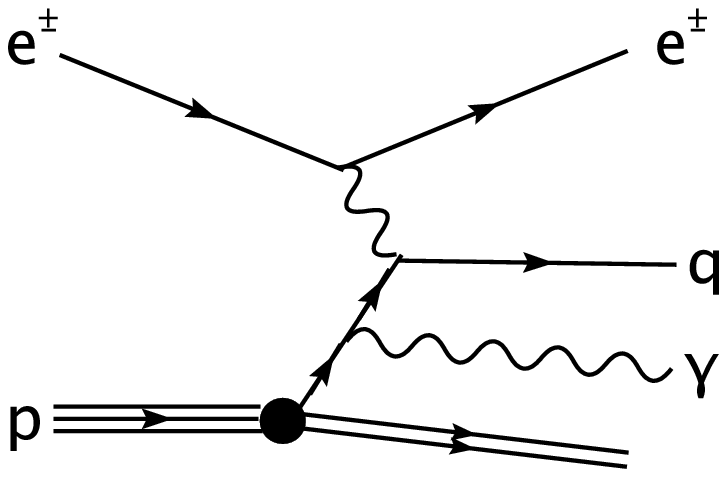}
    \hspace{0.04\textwidth}
    \includegraphics[width=0.30\textwidth]{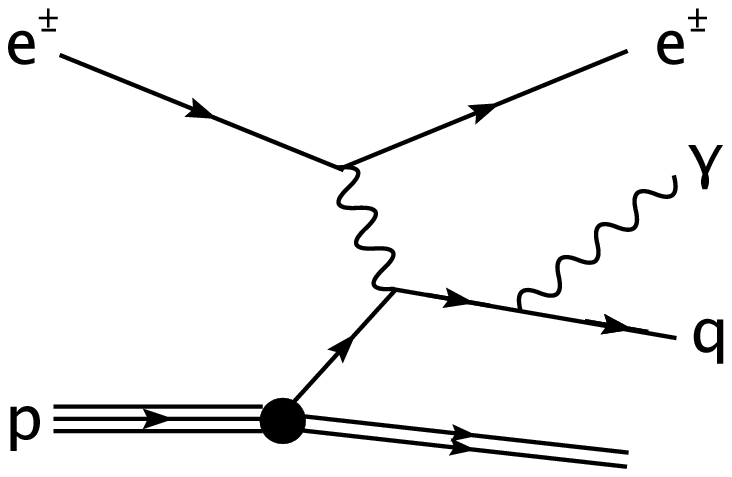}
 \end{center}
  \begin{picture}(0,0)
  \put(23,60){\textsf{\LL}}
  \put(23,25){\textsf{\QQ}}
  \end{picture}
    \caption{Leading order diagrams for isolated photon production in DIS.
The upper diagrams illustrate  isolated photon production by radiation from the electron line ($LL$), while the lower
diagrams correspond to  production via radiation from the quark ($QQ$, without the contribution from quark fragmentation).
    }
    \label{fig:feynman}
\end{figure}

The measurement of isolated photons in DIS
provides a test of perturbative QCD in a kinematic range with two hard scales: 
the transverse energy of the emitted photon $E_T^\gamma$ and $Q^2$. 
Isolated photons in DIS are produced at lowest order ($\alpha^3\alpha_s^0$)
 as shown in figure~\ref{fig:feynman}. 
Already at this order a  jet can be produced
         in the hadronic final state in addition to 
jets associated with the proton remnant,
 due to the electron or photon recoil. 

The final state photon may be emitted by a quark ($QQ$) and 
by wide angle radiation from the lepton ($LL$). 
 The interference contribution ($LQ$) is expected to be
   small.
Since the photon and the scattered electron are well separated in the
 present analysis, low angle QED radiation is suppressed.
The \QQ contribution has two different origins: the  direct radiation of a 
photon from the quark and  the  fragmentation of the quark into a jet containing a photon 
which carries a large fraction of the jet energy.
This quark-to-photon fragmentation contribution is
suppressed by the isolation requirement for the photon.

This paper presents a measurement of  isolated  photon production in DIS 
$e+p\rightarrow e+\gamma + X$.
Photons are identified using a multivariate
   analysis of the shapes of the calorimeter energy deposits to 
reduce the background from  neutral hadrons and their decay products. 
The photons are then used together with the other particles in the event, with the exception of the scattered electron, to reconstruct jets.
 The isolation of the photon is ensured by requiring that it carries at least 90\% of the transverse momentum of the jet containing the photon.
Isolated photons with  transverse energy $3 < E_T^\gamma < 10$~GeV
and  pseudorapidity   $-1.2 < \eta^\gamma < 1.8$
are selected in DIS events in the kinematic regime 
  $4<Q^2<150$~GeV$^2$, 
inelasticity $y>0.05$ and a mass of the hadronic system $W_X>50 $~GeV. 
The production of additional jets besides the photon jets in these events is also investigated.
The current analysis significantly extends the kinematic range probed by the ZEUS measurement~\cite{ZEUSdis}.
The results are compared to a recent leading order (LO), $\mathcal{O}(\alpha^3\alpha_s^0)$,
calculation~\cite{Thomas1,Thomas} and to  predictions of the Monte Carlo (MC) models 
PYTHIA~\cite{PYTHIA}, simulating the \QQ process,   and RAPGAP~\cite{RAPGAP} for the \LL process.
The cross sections for a photon plus at least one jet are further compared to a NLO calculation\cite{kramer}.
\section{H1 Detector}
\label{exp}
A detailed description of the H1 detector can be found in~\cite{H1det}. 
In the following, only  detector components relevant to this analysis are 
briefly discussed. The origin of the H1 coordinate system is the nominal $ep$ 
interaction point, with the direction of the proton
beam defining the positive $z$-axis (forward direction). 
Transverse momenta are measured in the $x$-$y$~plane. Polar ($\theta$) and azimuthal ($\phi$) angles are measured with respect to this reference system.
The  pseudorapidity is defined to be  $\eta = - \ln \tan(\theta/2)$.
\par
In the central region \mbox{($20^\circ\!<\!\theta\!<\!165^\circ$)} the interaction region is surrounded by the central tracking
system, which consists of a silicon vertex detector, drift chambers
and multi-wire proportional chambers,
all operated within a solenoidal magnetic
field of $1.16 \rm\ T$.
The trajectories of charged
particles are measured 
in the central tracker
with a transverse
momentum resolution of $\sigma(P_T)/P_T \simeq 0.005 \,
P_T \, /\mathrm{GeV} \oplus 0.015$.
The forward tracking detector  
and the
backward drift chamber (operated in 1999-2000) or
backward proportional chamber (for 2003-2005) 
measure
tracks of charged particles at smaller \mbox{($7^\circ\!<\!\theta\!<\!25^\circ$)} and larger \mbox{($155^\circ\!<\!\theta\!<\!175^\circ$)} polar angle
than the central tracker, respectively.
In each event the $ep$ interaction vertex is reconstructed from the measured charged tracks.
\par 
The liquid argon (LAr) sampling calorimeter~\cite{Andrieu:1993kh} surrounds the tracking
chambers. It has
 a 
 polar angle coverage of 
\mbox{$4^\circ\!<\!\theta\!<\!154^\circ$} and full azimuthal acceptance.
 It consists of an inner electromagnetic section with lead absorbers and an outer hadronic section with steel absorbers.
The calorimeter is divided into eight wheels  along the beam axis, 
each of them segmented in $\phi$ into eight modules, separated by small regions of inactive material.
The electromagnetic and the hadronic sections are highly segmented in the transverse and the longitudinal 
direction with about $44~000$ cells in total. The granularity is larger in 
the electromagnetic part and increasing in both sections in the forward direction.
For particles coming from   the $ep$ interaction region, 
the laterally projected cell size in the electromagnetic part varies between $5 \times 5$~cm$^2$ in the 
forward and at most $7 \times 13$~cm$^2$ in the central region. 
The longitudinal segmentation in the different wheels varies from   
three (central) to four (forward)
layers in the electromagnetic and from four to six in the hadronic section.
The first electromagnetic layer has a thickness of about $3$ to $6$ radiation lengths for particles coming from the interaction region.
Electromagnetic shower energies are measured with a precision of
$\sigma(E)/E=12\%/\sqrt{E/\gev}\oplus 1\%$ and hadronic energies with
$\sigma(E)/E=50\%/\sqrt{E/\gev}\oplus 2\%$, as determined in test beam experiments~\cite{Andrieu:1994yn,h1testbeam}. 
In the backward
region \mbox{$153^\circ\!<\!\theta\!<\!178^\circ$}, particle energies are measured
by a lead-scin\-tillating fibre spaghetti calorimeter (SpaCal)~\cite{Appuhn:1996na}. 
\par 
The luminosity is determined from the rate of the  Bethe-Heitler process 
$ep\!\rightarrow\!ep\gamma$, measured using a photon detector located close 
to the beam pipe at $z=-103$~m,  in the backward direction. 
\par
DIS events at $Q^2$ values up to $150$~GeV$^2$   are triggered by the energy deposition of the scattered electron in the SpaCal. 
For events with the scattered electron entering the SpaCal at low radii, 
additional trigger signals are required from the central drift chambers~\cite{dcrphi,ftt} and  from the central proportional chambers~\cite{cip,eichi,cip2000}.
 The trigger efficiency for DIS events containing an electron in the Spacal angular acceptance with  an energy 
above $10$~GeV is greater than $98$\%.
\section{Monte Carlo Simulations}
Monte Carlo simulations are used to
correct the data for detector acceptances, inefficiencies
and migrations and to compare the measured cross sections
with MC model predictions.
\par
The two generators PYTHIA~\cite{PYTHIA} and  RAPGAP~\cite{RAPGAP} 
are used to generate  events with photons produced in the hard interaction.
PYTHIA simulates the contribution of photons radiated from the struck quark ($QQ$).
The contribution of photons radiated by the electron ($LL$) is simulated 
using RAPGAP and 
denoted ``RAPGAP rad.'' in the following.
The small contribution from interference~\cite{Thomas1} is neglected.
 Both generators 
calculate the hard partonic interaction in LO QCD ${\cal O}(\alpha^3\alpha_s^0)$.
Higher order QCD radiation is modelled using initial and final state parton showers in the leading log approximation~\cite{Bengtsson:1987rw}. 
The fragmentation into hadrons is simulated  using the LUND string model~\cite{lund} as implemented in JETSET\cite{jetset}. The simulations use the CTEQ6L proton parton densities~\cite{Stump:2003yu}.

The measurements presented in this paper show that the data is well 
described by the two MC contributions if 
PYTHIA is scaled by  a factor $2.3$ and RAPGAP is not scaled.
This combined ``scaled signal MC''  is used to correct the data, 
whereas the unscaled MC
prediction (``signal MC'') is compared to the cross section measurements.

As an alternative, the HERWIG~\cite{HERWIG} generator is used to model the  \QQ contribution.
HERWIG simulates the fragmentation into hadrons  through the decay of colourless parton clusters and uses the
equivalent-photon approximation for the incoming photon beam. 
Isolated photon production in DIS is derived approximately as Compton scattering between the photon and a quark. 
This approximation is not valid for $Q^2$ above a few GeV$^2$, therefore HERWIG is only used to estimate the systematic uncertainties due to the fragmentation model.

The main SM background  is due to    photons produced in hadron decays in DIS events.
It is modelled using the RAPGAP generator, with initial and final state radiation  switched off.
This contribution is denoted by ``RAPGAP non-rad'' in the following.

The multivariate shower shape analysis used to identify the photons requires
 high statistics
samples of shower simulations in the whole  phase
space of energy and pseudorapidity.
Samples of events containing single particles are simulated. 
In each sample, corresponding to single 
photons or single neutral hadrons decaying to photons,  the particles
 are uniformly generated in 
 pseudorapidity and energy. 
These samples are generically named ``single particle samples'' in the 
following.

All generated events are passed through  a full GEANT~\cite{geant} simulation of the 
H1 detector and through the same reconstruction and analysis programs as used for the data.
\section{Event Selection}
The event sample used in this analysis was collected with the H1 detector
at HERA in the period  $1999$ to $2005$ at a centre-of-mass energy of $319$~GeV.
The  corresponding integrated luminosity is $227$~pb$^{-1}$.
In a first step, DIS events are selected with the scattered 
electron\footnote{The analysis uses data from periods when the beam lepton was either a positron or an electron. 
Unless  otherwise stated,  the term electron  refers to both the electron and the positron.} 
measured in the SpaCal.
In a second step, a subsample of DIS events with an isolated photon candidate in the LAr calorimeter is selected. 

\subsection{Selection of DIS events}

DIS events are selected with  the scattered electron identified in the SpaCal as a
            compact electromagnetic cluster~\cite{elan} with an energy
$E_e>10$~GeV and a polar angle $\theta_e<177^{\circ}$.
Matching signals in the backward tracking chambers are required for electron candidates with  $E_e<18$~GeV.
 The scattering angle of the electron is determined from the measured impact 
position in the backward tracking chamber, the position of the energy cluster in the SpaCal and the 
reconstructed primary vertex. 

\par
Background from   events at low $Q^2$, in which the electron escapes through the beam pipe and a hadron fakes the electron signature, 
is suppressed
 by the requirement that the difference $\Sigma(E-p_z)$ between the total energy 
 and the longitudinal momentum be in the range  
\mbox{$35\!<\!\Sigma(E-p_z)\!<\!70$}~GeV,
 where the sum includes
 all measured hadronic final state particles and the scattered electron.

Non-$ep$ background is removed by restricting the $z$-coordinate of the  event vertex  to be within $\pm 40$~cm of the average  vertex position and by requiring at least one good track in 
the central tracking system with the polar angle $30^{\circ}<\theta<150^{\circ}$ and not associated to the electron. 

The energy $E_e$ and polar angle $\theta_e$ of the scattered electron candidate
are used to reconstruct  $y$ and  $Q^2$ according to 
\mbox{$Q^2\!=\!2\,E^0_e E_e\!(1 + \cos\theta_{e})$} and 
\mbox{$y\!=\!1-E_e (1-\cos\theta_e)/(2E^0_e)$}, where $E^0_e$ is the electron beam energy.
The events are selected in the kinematic region  \mbox{$4\!<\!Q^2\!<\!150$~GeV$^2$} and $y >0.05$.
\subsection{Selection of isolated photon candidates and jets}\label{sel_jet}

Photon candidates are identified as 
clusters in the electromagnetic section of the LAr calorimeter
 with a transverse energy  $3 < E_T^\gamma < 10 $~GeV and pseudorapidity $-1.2 < \eta^\gamma < 1.8$ in the H1 laboratory frame. 
 The candidates are rejected if they are  close to  
inactive regions between calorimeter modules~\cite{Andrieu:1993kh} 
 or if a track geometrically matches
the electromagnetic cluster with a distance of closest approach to the cluster's barycentre of 
less than $20$~cm. 
Neutral hadrons that decay into multiple photons, predominantly $\pi^0 
\rightarrow \gamma\gamma$, constitute the main background. 
In most cases such decay photons
  are merged into one electromagnetic cluster, which tends to have a wider
 transverse
  distribution than that of a single photon. 
The transverse radius\footnote{For a definition of the transverse radius see section~\ref{sec:photon}.}  $R_T$ of the
  photon candidate cluster is therefore required to be smaller than $6$~cm.
  In addition, the invariant mass of the cluster, when combined with the closest neighbouring 
electromagnetic
   cluster with an energy above $80$~\mev, must be larger than $300 \mev$. This
  requirement rejects candidates that originate from $\pi^0$~decays with
  two photons reconstructed in separate clusters.
Only events with exactly one photon candidate are accepted.
Less than $1$\% of the events are rejected because more than one photon 
candidate is found. 

The mass of the final state hadronic system is reconstructed from the four-momenta of the incoming electron ($p_e$) and proton ($p_p$), the scattered electron ($p_e^\prime$) and the 
photon candidate ($p_\gamma$) as $W_X=\sqrt{(p_e+p_p-p_e^\prime - p_\gamma)^2}$.
The contribution  from elastic Compton scattering \linebreak ($p+e \rightarrow p+e+\gamma$) is 
suppressed by requiring
$W_X>50$~GeV. 
\par
Final state hadrons  are reconstructed from deposits in the LAr 
calorimeter in combination with tracking information. 
Following  the so-called ``democratic" procedure~\cite{Glover:1993xc,Buskulic:1995au},  
the photon candidate and  the 
reconstructed hadrons in each event are combined into massless jets using the 
 $k_T$ algorithm~\cite{jetalgo}.
The algorithm is used with a $P_T$-weighted recombination scheme
and with the separation parameter $R_0$  set to $1$.
 Jets are selected with a  transverse momentum of
$P_{T}^{jet} > 2.5 $~GeV and a pseudorapidity in the range $-2.0<\eta^{jet}<2.1$.
Due to the harder kinematical cuts for the photon candidate there is always a jet containing the photon candidate, called the photon-jet. 
 All other jets are classified as hadronic jets.
For hadronic jets the $\eta^{jet}$-range is restricted to $-1.0<\eta^{jet}<2.1$. According to the MC 
simulation, the hadronic jets are well correlated to the partonic jets even at low transverse energies.
To ensure isolation of the photon, the  
fraction $z$ of the transverse energy of the photon-jet carried by the photon candidate has to  be larger than $90$\%.
This definition of the isolation requirement is
   stable against infrared divergences and thus well suited for comparisons
   with perturbative QCD calculations.
The isolation requirement largely suppresses background from
photons produced in hadron decays.
\par
\begin{figure}[Hhh]
    \includegraphics[width=0.473\textwidth]{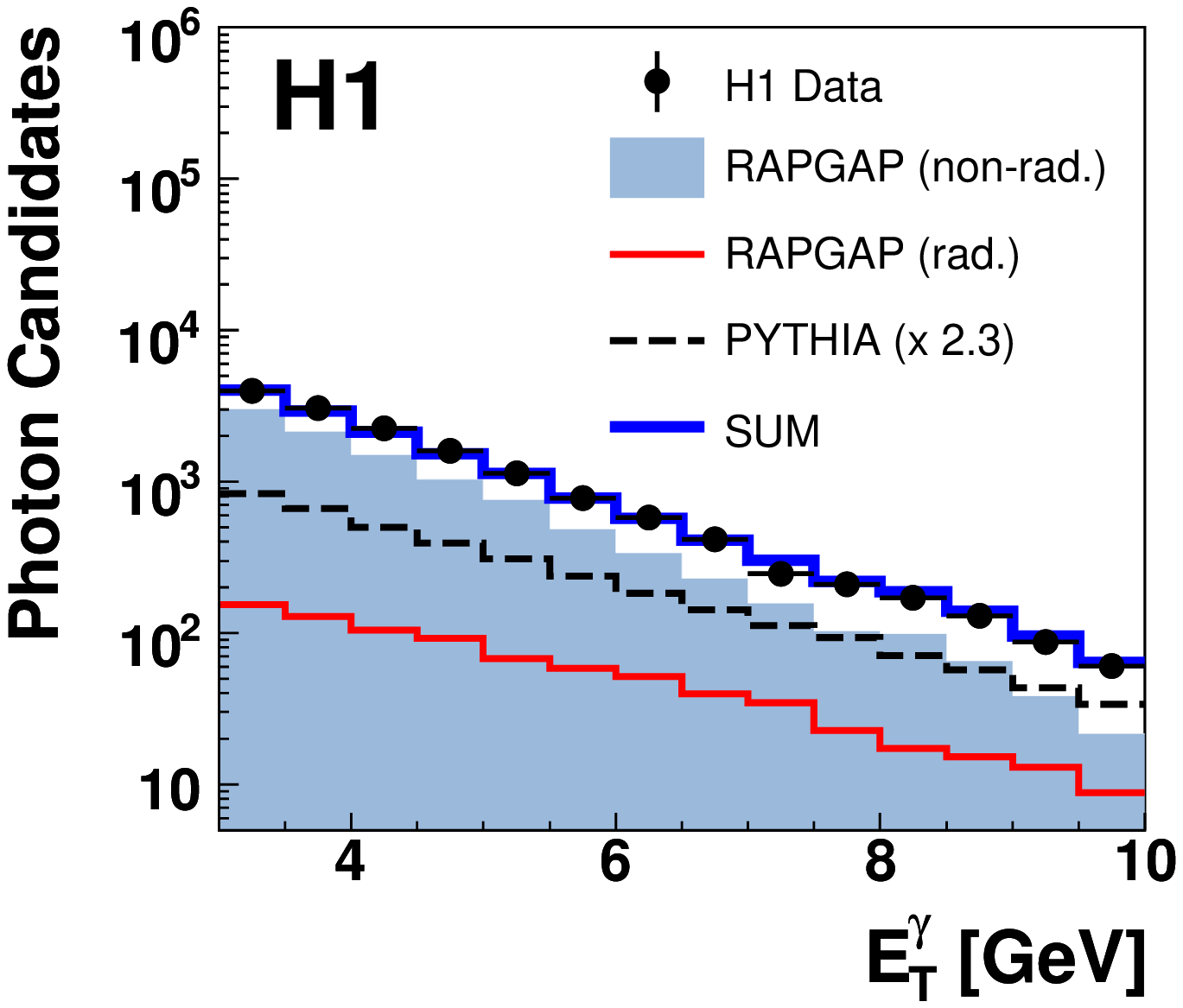}
    \hspace{0.05\textwidth}
    \includegraphics[width=0.473\textwidth]{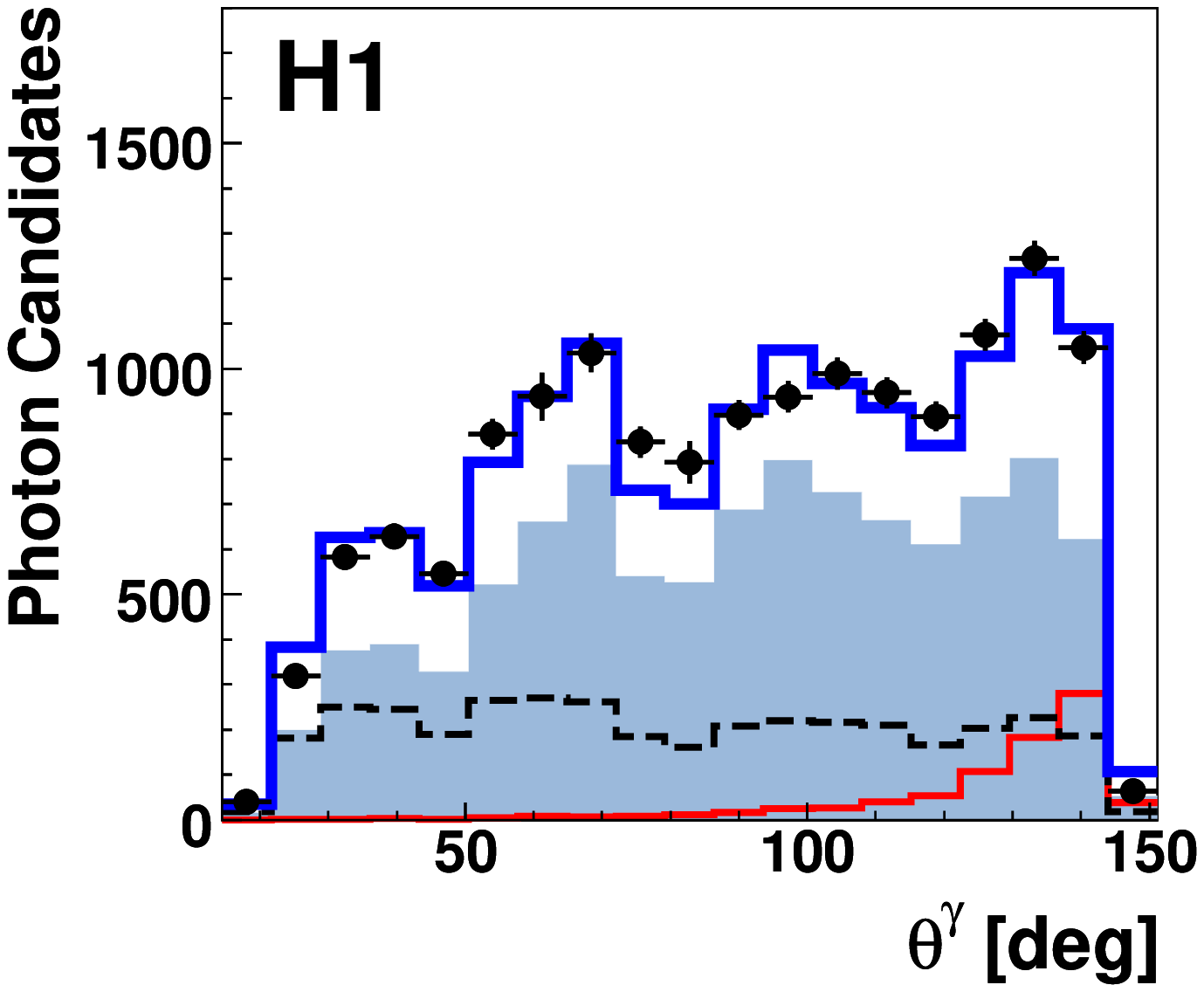}
  \begin{center}
    \caption{
Distributions of (a) $E_T^\gamma$ and (b) $\theta^\gamma$ for isolated  photon
candidates in the final event sample.
Data are shown as points with error bars. 
The bold solid histogram shows the sum of the
expectation from RAPGAP~(non-rad.) for neutral hadron background (shaded), from
PYTHIA for radiation from the quark scaled by a factor of 2.3 (dashed
line) and from RAPGAP~(rad.) for radiation from the electron (solid line). 
The unshaded area corresponds to the estimated isolated photon contribution (RAPGAP(rad) plus PYTHIA$\times$2.3).}
    \label{fig:clusterplots}
  \end{center}
  \begin{picture}(0,0)
  \put(66,106.5){\textsf{(a)}}
  \put(150.2,106.5){\textsf{(b)}}
  \end{picture}
\end{figure}

The distributions of the transverse energy and of the polar angle of the isolated photon
 candidates  are shown in figure~\ref{fig:clusterplots} together with the \mc predictions for the scaled signal and the background. The sum of the MCs describes the data well.
\par
\begin{figure}[Hhh]
    \includegraphics[width=0.473\textwidth]{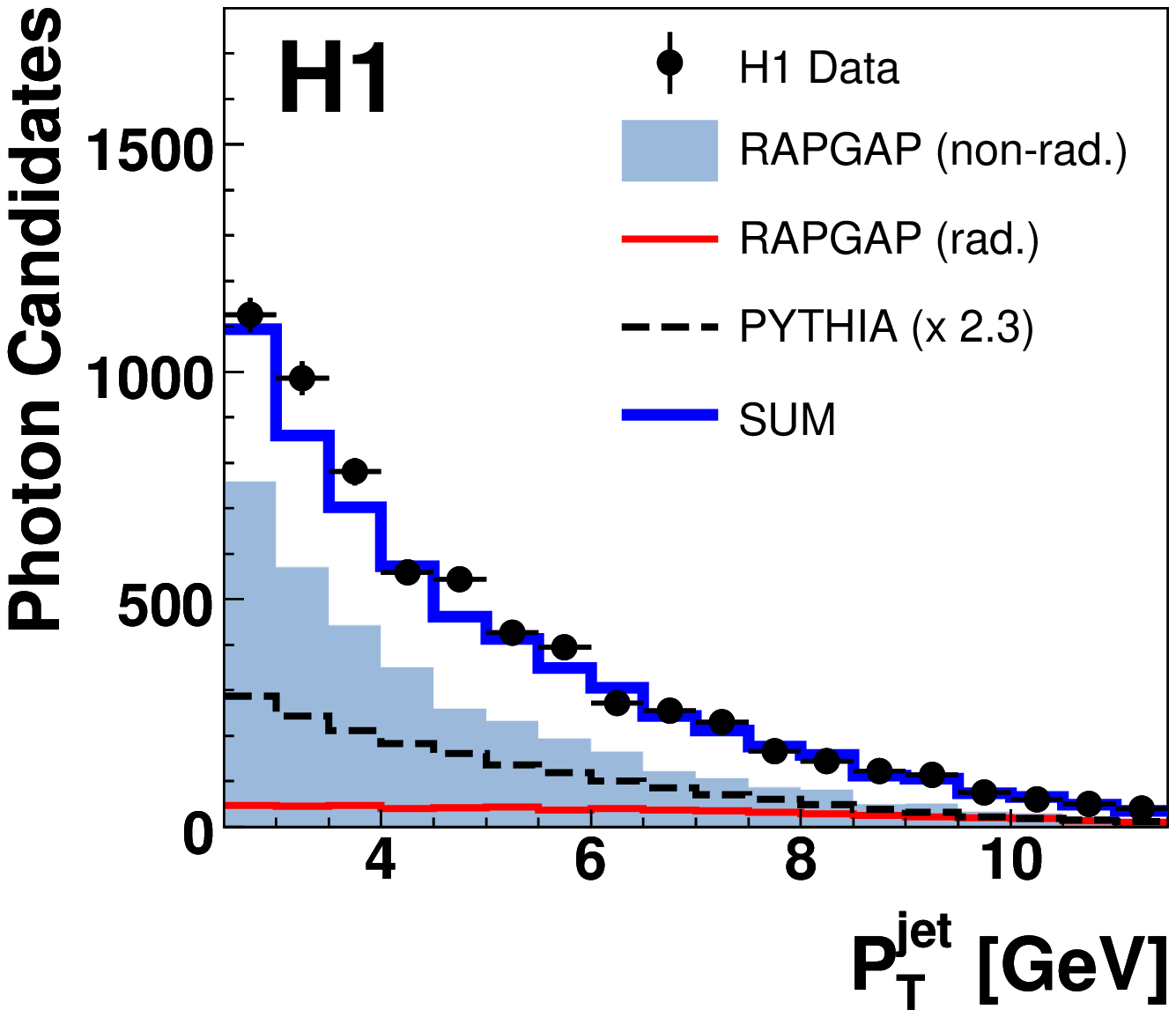}
    \hspace{0.04\textwidth}
    \includegraphics[width=0.473\textwidth]{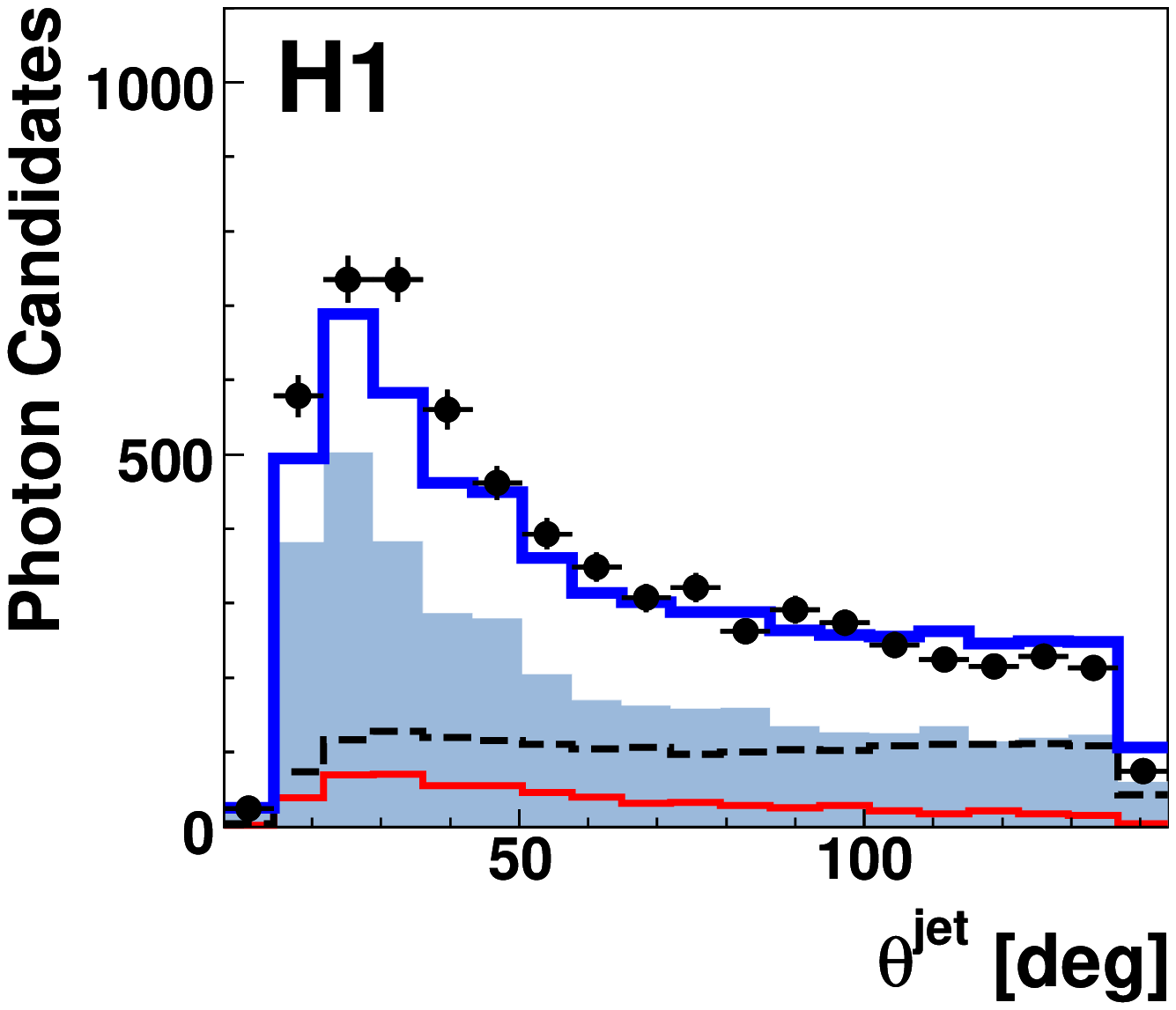}
  \begin{center}
    \caption{
Distributions of (a) the transverse momentum and (b) the polar angle of
the hadronic jet with the highest transverse momentum in events with an isolated photon candidate. 
The data are compared to the MC predictions, as described in the caption of
 figure~\ref{fig:clusterplots}.}
    \label{fig:jetplots}
  \end{center}
  \begin{picture}(0,0)
  \put(66,91.0){\textsf{(a)}}
  \put(150.2,91.0){\textsf{(b)}}
  \end{picture}
\end{figure}

 
 The samples of events with either no hadronic jet  or at least one hadronic jet 
  are called ``photon plus no-jets'' and ``photon plus jet'', respectively.
The $P_T^{jet}$ and $\theta^{jet}$  distributions for the hadronic jet with the largest transverse momentum are shown in figure~\ref{fig:jetplots}. Both distributions are reasonably well described by the sum of the  scaled signal and background MCs.

A total of $14670$ events with a scattered electron and an isolated photon candidate are selected, 
 of which  $6495$ have at least  one additional hadronic jet.


\section{Photon Signal Extraction}\label{sec:photon}

\subsection{Shower shape analysis}

The extraction of the photon signal  exploits the fine granularity 
of the electromagnetic part of the LAr calorimeter. 
In order to discriminate between signal photons and the background from
neutral hadrons and their decay products, the calorimeter cluster
corresponding to the isolated photon candidate is further analysed using
the following six shower shape variables calculated from the measurements of the individual cells composing the 
cluster~\cite{carsten}:

\begin{enumerate}
\item The fraction of the energy of the electromagnetic cluster contained in the 
cell with the largest energy deposit (``hottest cell''). 
\item The fraction of the energy of the electromagnetic cluster contained in four or eight 
(depending on the granularity of the calorimeter) contiguous cells in the first two calorimeter layers.
 The cells include the hottest cell and are 
chosen to maximise the energy which they contain  (``hot core'').  
This and the first variable are sensitive to the compactness 
of the cluster in the calorimeter.
The values of these variables are on average larger for photons than for the background.
\item The fraction of the cluster's  energy detected in the first calorimeter layer (``layer 1''), which
is expected to  be larger on average for multi-photon clusters than 
for those initiated by a single photon. 
\item The transverse\footnote{In the context of the cluster shape analysis the transverse plane is defined as perpendicular
  to the direction of the photon candidate.}  symmetry $S_T$ of a cluster defined as the ratio of 
the spread (defined by root mean squared) of the transverse
 cell distributions along the two principal axes.
A photon cluster is  expected to be
  symmetric  with $S_T$ values close to unity, whereas multi-photon clusters are  typically more asymmetric and yield lower $S_T$ values.
\item The transverse radius of the cluster  defined as the square root of the second central transverse moment  $R_T=\sqrt{\mu_2}$, where the $k$'th central transverse moment of the cells distribution is given by
$\mu_k= \langle |\vec{r}- \langle\vec{r}\rangle |^k \rangle $. Here
 $\vec{r}$ is the transverse projection of a cell 
position and 
$\langle \vec{r} \rangle = (\sum_{i=0}^n E_i \vec{r_i})/\sum E_i$  the 
energy weighted average of the cell 
positions $\vec{r_i}$ in the plane transverse to the photon direction.
As explained in section 4.2, only events with a cluster candidate of small transverse radius $R_T<6$~cm are selected for the 
multivariate analysis.
\item The transverse kurtosis $K_T$ is defined as
$K_T=\mu_4/(\mu_2)^2 - 3$.
It specifies how strongly the energy distribution is peaked and
is equal to zero for a Gaussian distribution.
\end{enumerate}
The discrimination power of signal and background becomes weaker at high transverse energies, 
where the multi-photon clusters become more similar to a single photon cluster. 
Therefore events with $E_T^\gamma>10$~GeV are excluded from the measurement, as described in section 4.2.

\begin{figure}[hhh]
    \includegraphics[width=0.473\textwidth]{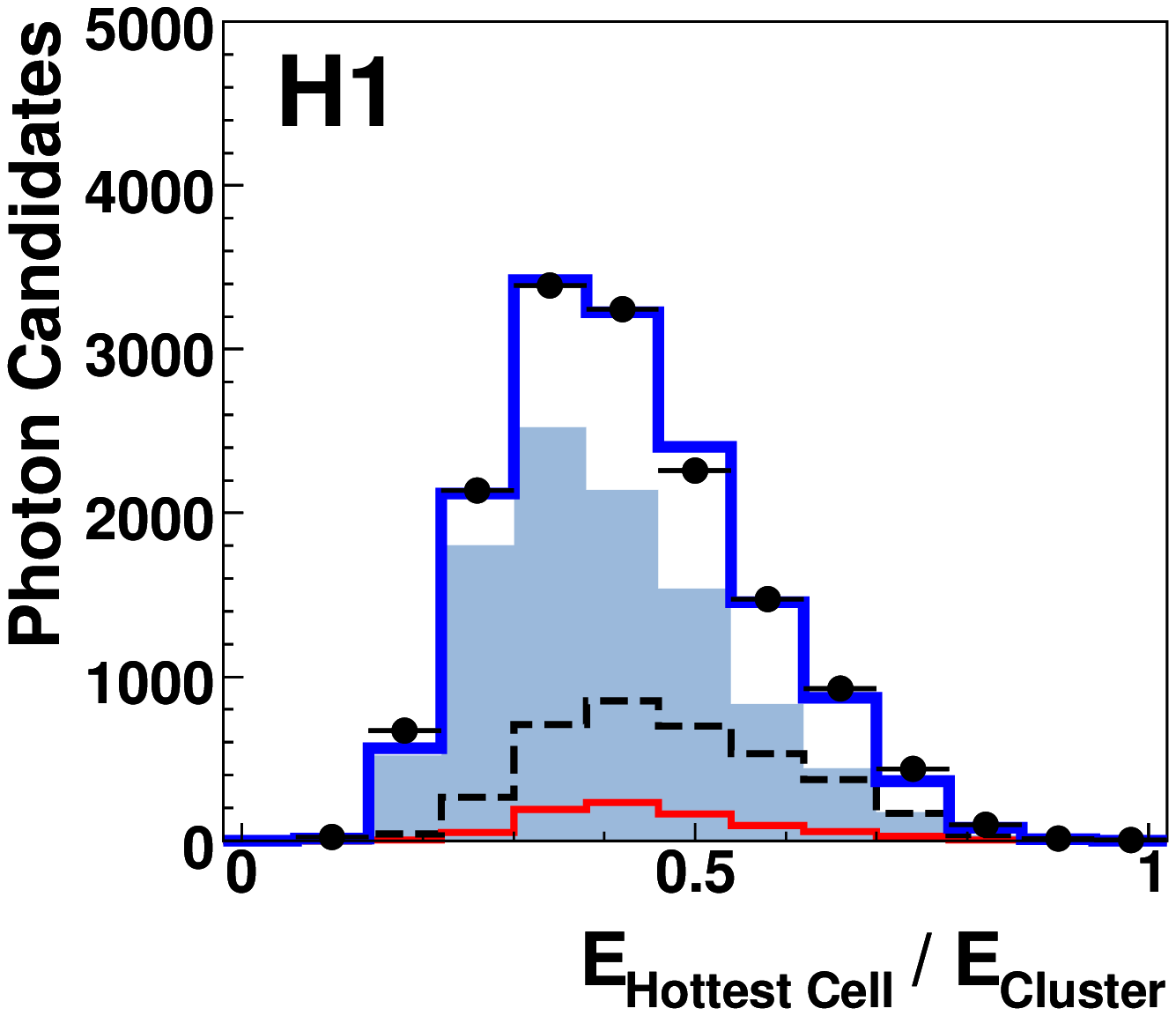}
    \vspace{0.3cm}
    \hspace{0.04\textwidth}
    \includegraphics[width=0.473\textwidth]{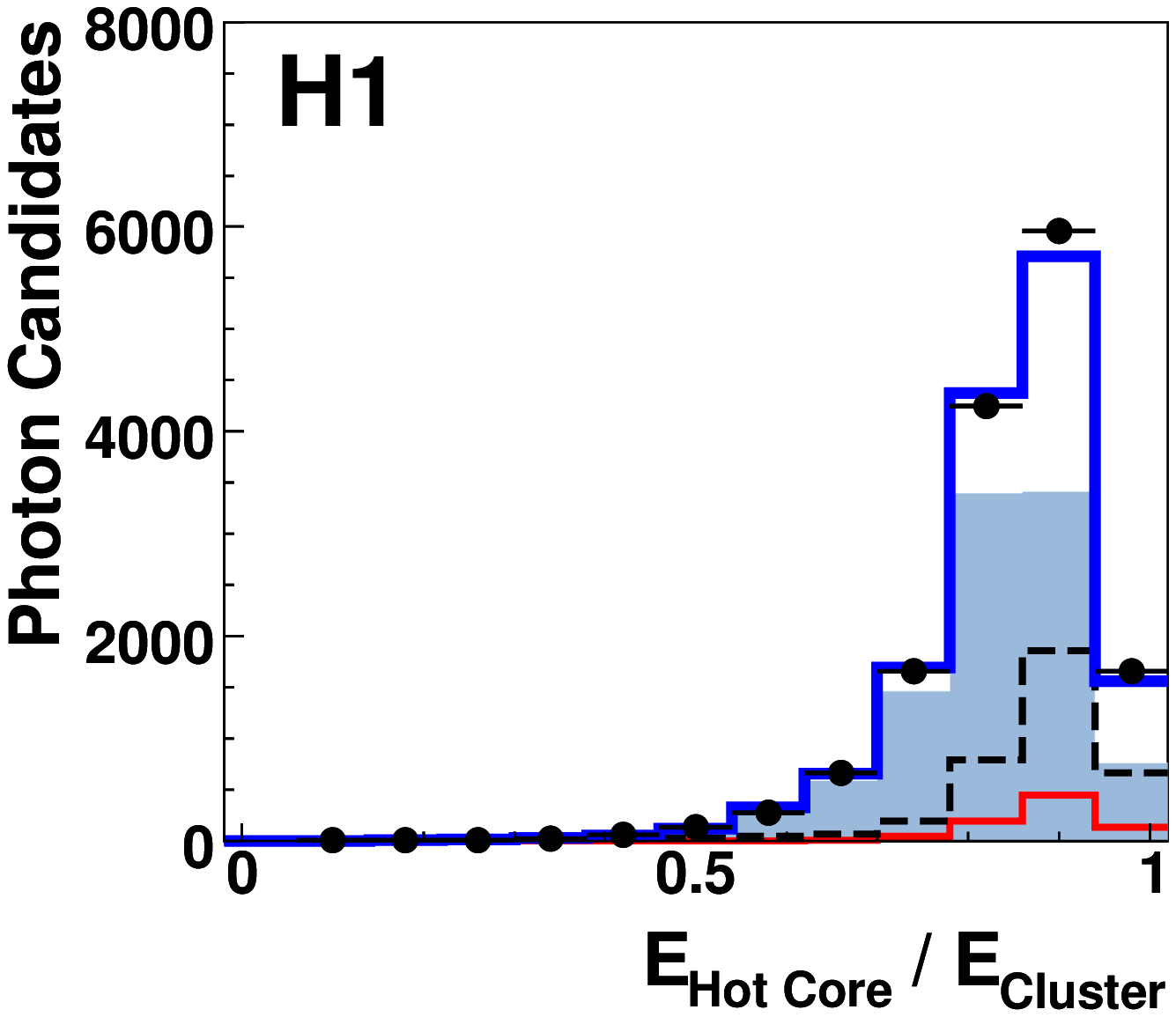}
    \includegraphics[width=0.473\textwidth]{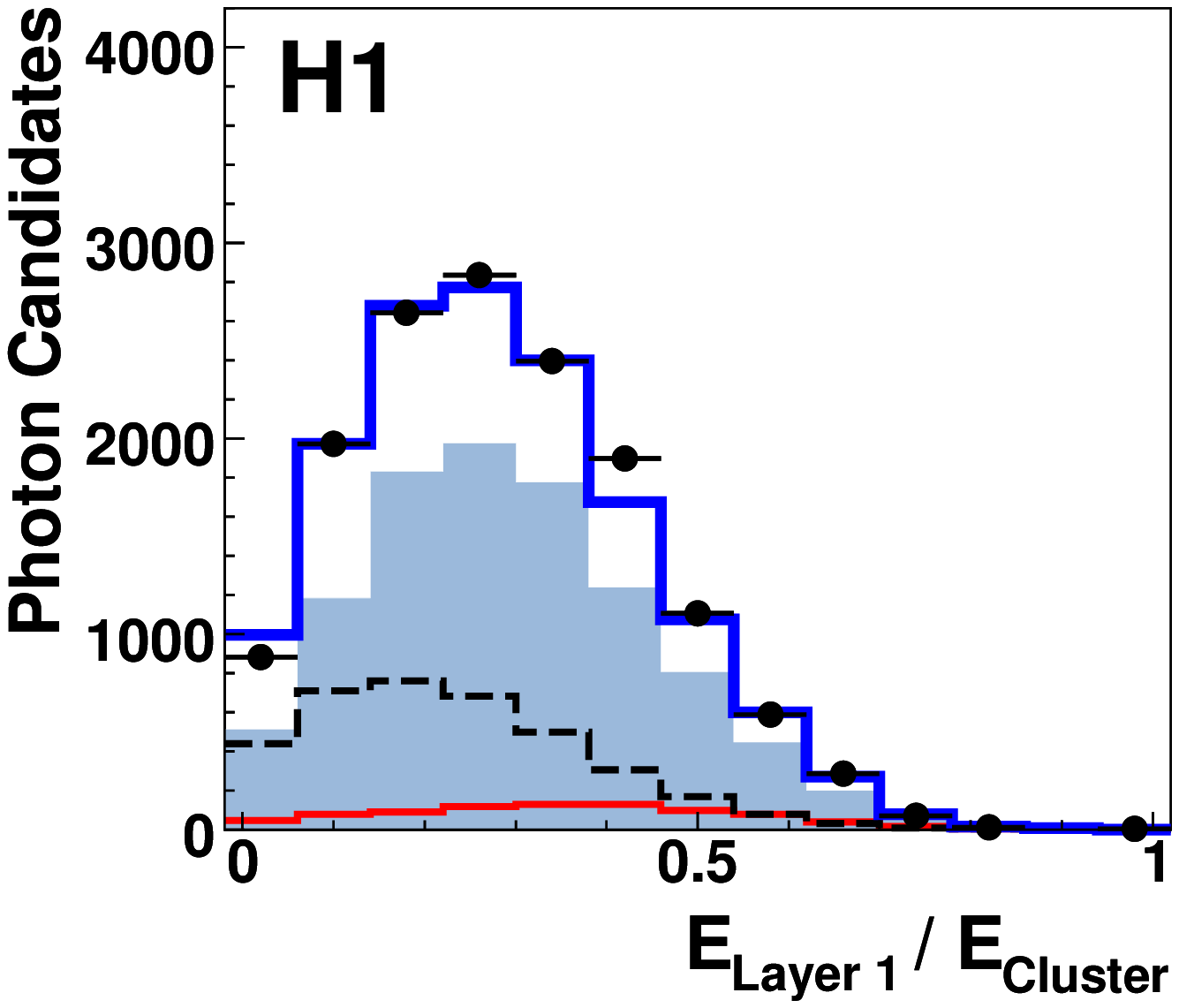}
    \vspace{0.3cm}
    \hspace{0.04\textwidth}
    \includegraphics[width=0.473\textwidth]{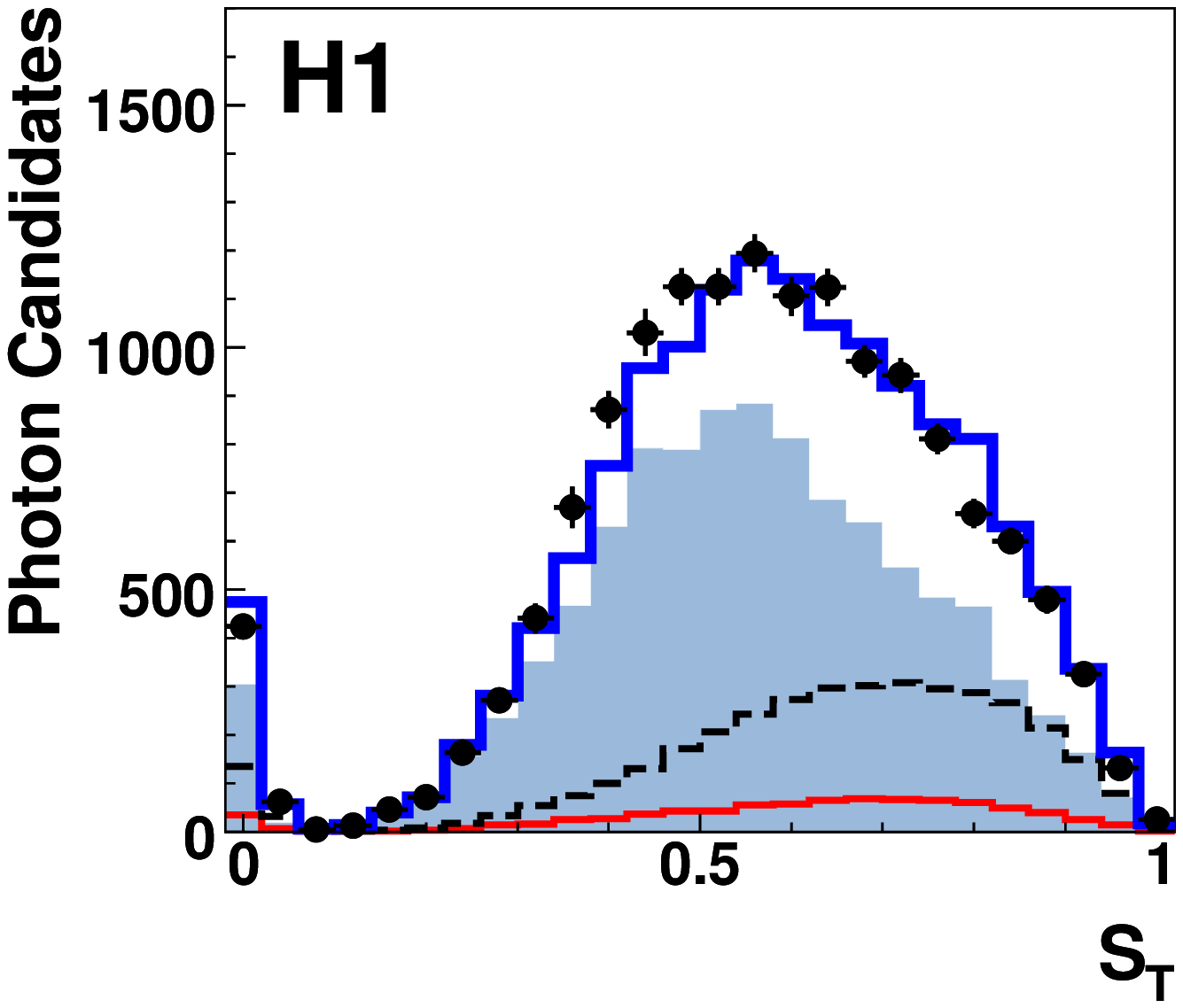}
    \includegraphics[width=0.473\textwidth]{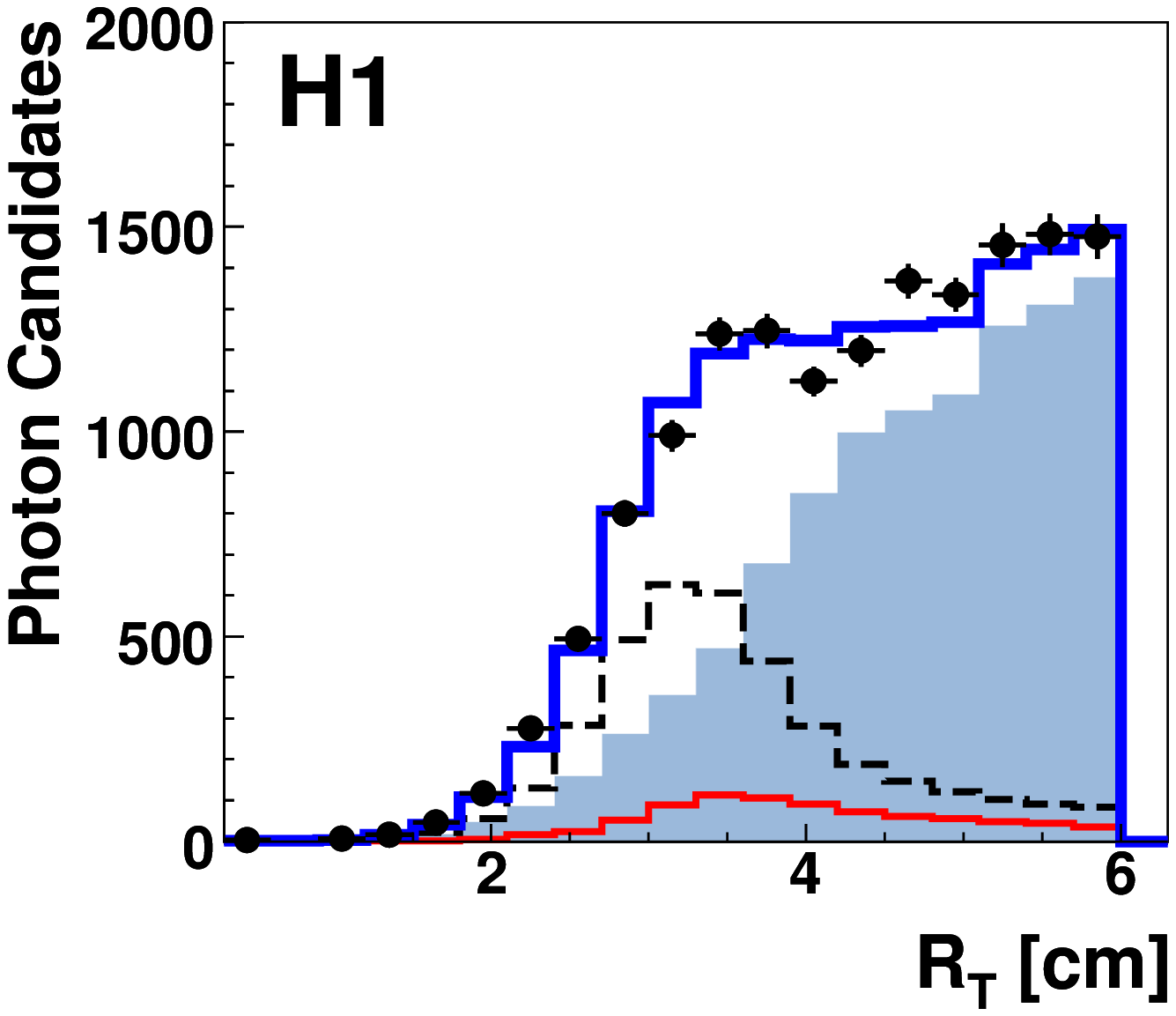}
    \hspace{0.04\textwidth}
    \includegraphics[width=0.473\textwidth]{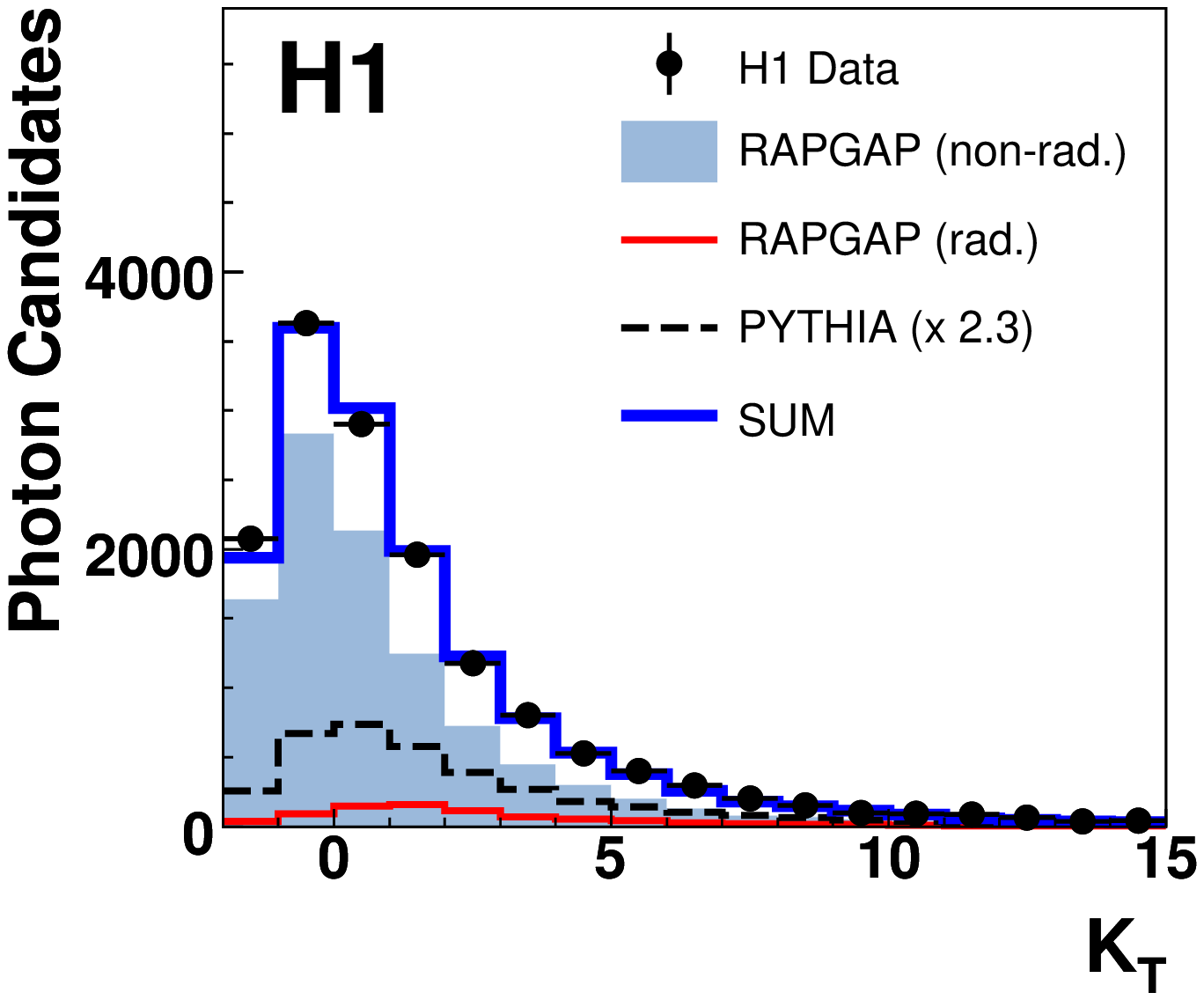}
  \begin{center}
    \caption{Distributions of the six  variables that are used to define the
discriminant for isolated photon identification: 
 (a) hottest cell fraction,
(b) fraction of the hot core, (c) first layer fraction, 
(d) transverse symmetry $S_T$, (e) transverse radius $R_T$ and 
(f) transverse kurtosis $K_T$.  
The data are shown with the MC predictions described in the caption to
figure~\ref{fig:clusterplots}. The shape difference between RAPGAP~(rad.) and PYTHIA arises 
from the different  distributions in phase space (see figure~~\ref{fig:clusterplots}b).}
    \label{fig:showershapes}
  \end{center}
  \begin{picture}(0,0)
    \put(66,237.0){\textsf{(a)}}
    \put(150.5,237.0){\textsf{(b)}}
    \put(66,171.5){\textsf{(c)}}
    \put(150.5,171.5){\textsf{(d)}}
    \put(66,106.0){\textsf{(e)}}
  \put(150.5,106.0){\textsf{(f)}}
  \end{picture}
\end{figure}

The distributions of the six shower shape variables are shown for the isolated
        photon candidates in figure~\ref{fig:showershapes}.
The data are compared with the sum of the background and the scaled signal  \mc 
distributions. A good agreement is observed.

\subsection{Signal extraction}
In order to discriminate between  single photons (signal) and single neutral hadrons (background), probability density functions   $p^{i=1,6}_{\gamma,bg}$ are 
determined for the six shower shape variables, using simulated ``single
      particles events'', described in section 3. 
The signal  probability density functions  $p^i_{\gamma}$ are simulated using single photon 
events whereas the background probability density functions $p^i_{bg}$ are approximated using a sample of events containing single neutral 
hadrons ($\pi^0$, $\eta$,
 $\eta^\prime$, $\rho$, $\omega$, $K^\star$, $K_L^0$, $K_S^0$, $n$ and
 $\bar{n}$). 
The relative contributions of the various neutral hadrons species 
are taken as predicted by the RAPGAP generator. 
In particular, $\pi^0$ and $\eta$ mesons contribute to 90\% of the background.


 An overlap of clusters of different particles  can occur due to large multiplicities specific to 
the hadronic environment in DIS. 
The overlap affects the photon candidate cluster shape. It is found to be 
important only for the background and leads
to a loss of photon candidates due to the distortion of the  
transverse cluster radius, which 
then exceeds in most cases the upper limit of $6$~cm required in 
the analysis (cf.~section 4.2). 
This loss due to
cluster overlap is modelled by supressing the background single particles according to a  probability $p_{co}$ proportional to  the transverse area of the cluster
$p_{co}= a \cdot R_T^2$ with $a=0.004$~cm$^{-2}$.
 The constant $a$ is determined by a comparison of single particles event samples with full MC simulation
 in phase space regions where sufficient statistics are available.
\par
\begin{figure}[hhh]
  \begin{center}
    \includegraphics[width=0.6\textwidth]{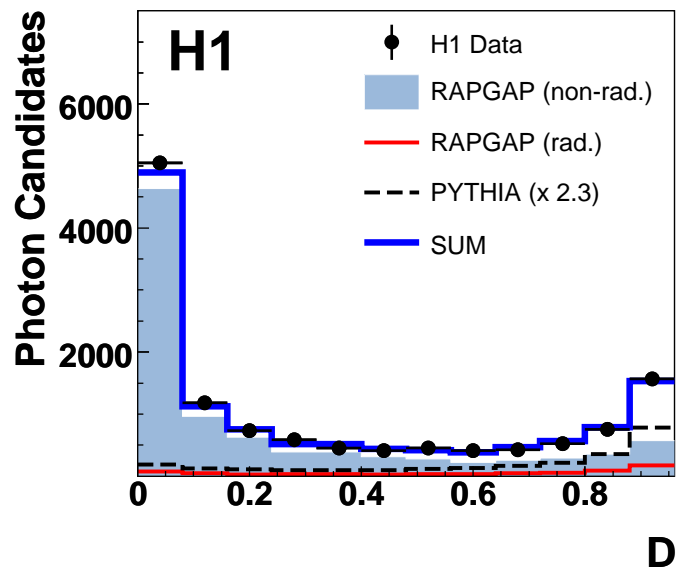}
    \caption{
The distribution of the discriminator ($D$) used in the identification of
isolated photon candidates for events that have passed the event
selection. The data are compared to the MC predictions described in the caption of
 figure~\ref{fig:clusterplots}.} 
    \label{fig:likelihood}
  \end{center}
\end{figure}

The multi-dimensional photon and background probability densities are
 taken as the product of the respective shower shape densities $P_{\gamma, bg}=\prod_{i=1,6} p^i_{\gamma,bg}$. 
For each event a discriminator ($D$) is formed, which is
 defined as the photon probability density divided by the sum of
 the probability densities for photons and background
$D=P_\gamma/(P_\gamma+P_{\mathrm bg})$.
The discriminator has in general larger values for isolated photons than for
the decay photons. 
 Figure~\ref{fig:likelihood}
 shows the discriminator distribution for  the data together with the predictions of the  background and the scaled signal MC. 
 The data are well described by the sum of the MC predictions.
\par
Since the shower shape densities 
vary significantly as a function of the cluster energy and depend on the granularity of the LAr calorimeter, the 
discriminator is determined in bins of ($E_T^{\gamma}$, $\eta^{\gamma}$), 
with three intervals in $E_T^{\gamma}$ and five in 
$\eta^{\gamma}$. 
The intervals in $\eta^{\gamma}$ correspond to the different wheels of the
 calorimeter.
The contributions of photons and neutral hadrons in any of the $15$  analysis bins is
 determined by independent minimum-$\chi^2$ fits 
 to the data discriminator distributions.
In each ($E_T^{\gamma}$, $\eta^{\gamma}$) bin, the $\chi^2$ function is defined to be

\begin{displaymath}
  \chi^{2}=\sum_{i}
  \frac{{\left(
        N_{data,i}-N_{bg}d_{bg,i}-N_{sig}d_{sig,i} \right)}^{2}}
  {\sigma^{2}_{data,i} + N_{bg}^{2}\,\sigma^{2}_{bg,i} +
    N_{sig}^{2}\,\sigma^{2}_{sig,i} }
\end{displaymath}

\noindent where the sum  runs over the bins of the discriminator distributions.
$N_{data,i}$ is the number of data events in the $i$'th bin.
$d_{sig,i}$ and $d_{bg,i}$ denote the $i$'th bin content of the
  signal and background
  discriminator distribution, respectively, normalised to unity.
The  $\sigma_{i}$ represent the associated statistical errors.
$\!N_{sig}$ and $N_{bg}$, represent the number of signal  and background events respectively, and are 
  determined by the fit.
If the content in any data histogram bin is small ($N_{data,i}<7$), adjacent bins are merged.
The fit is performed separately for the three samples selected as presented 
in section 4.2: inclusive, photon plus no-jets and photon plus jet.
The fit quality is acceptable in all differential bins.

The total number of isolated photons is obtained by summing the $N_{sig}$ from all analysis bins.
As a result $4372 \pm 145$ signal events are attributed to the inclusive data sample. In the
   \nojet and \jets subsamples $1755 \pm 106$
   and $2606 \pm 95$ signal events are found, respectively.
  The signal fraction is $29$\% in the  inclusive sample, 
$20$\% for the \nojet sample and $39$\% for the \jets sample. 

\subsection{Cross section determination}

In each bin of the kinematic variables, the cross section $\sigma$ is computed 
from the number of events with photons in the corresponding bin  as $\sigma = N_{sig}/(\mathcal{L}\cdot \epsilon)$, where  $\mathcal{L}$ is the luminosity. 
The correction factor $\epsilon$ takes into account the acceptance, trigger and reconstruction efficiencies, and migration between the bins.
It is calculated using the scaled signal MC.
Bin averaged cross sections are quoted in all tables and figures.
The total inclusive cross section is obtained by summing the 
measured cross sections from  all 15 analysis bins.
The bin averaged single differential cross sections
$d\sigma/dE_T^\gamma$ and $d\sigma/d\eta^\gamma$  are obtained accordingly by summing all corresponding bins in $\eta^\gamma$ and $E_T^\gamma$, respectively. 
The single differential cross section $d\sigma/dQ^2$ is determined by fitting the  discriminator distributions separately in five different bins in $Q^2$. 
In these fits, the signal and background
discriminator distributions in each $E_T^\gamma$ and $\eta^\gamma$ bin are assumed to be independent of $Q^2$.  
It has been verified that the variation of the $Q^2$ dependence have indeed a negligible effect.

\section{Systematic Uncertainties}
Two additional event samples 
are used for the determination of systematic errors and in-situ energy calibration. 
The first  sample, containing Bethe Heitler events,  $ep\rightarrow e\gamma p$, consists of events with an electron reconstructed 
in the LAr 
calorimeter, a photon in the SpaCal and nothing else in the detector. 
The second, complementary,  sample is 
selected by requiring  an electron in the SpaCal and a photon in the LAr calorimeter  and no other particle detected in the event. 
Such events originate to a large part from deeply virtual Compton scattering  $ep\rightarrow e\gamma p$.
These independent event selections, denoted BH and DVCS respectively,
provide a clean sample of electromagnetic clusters in the LAr 
calorimeter. 
\par
The effects of the different  systematic errors on the cross section  are 
evaluated by applying  variations to the \mc simulation.
The following uncertainties are considered:
\begin{itemize}
\item The measured shower shape variables in 
the  BH and DVCS event samples  are compared to simulated single particle photons and electrons, respectively. 
The uncertainty on the shower shape simulation is estimated by distorting the discriminating 
variables within the limits deduced from the differences observed for the control samples (BH and 
DVCS) between data and simulation.
The fits for the signal extraction (section 5.2) are  repeated with the distorted distributions of the discriminating variables.
The resulting systematic error on the total inclusive cross section is $+10.2$\%   and $-12.8$\%. 
It varies between $11$\% and $25$\% for
 the single differential cross sections.
 The error increases with increasing $E_T^\gamma$ and towards large $\eta^\gamma$ and is independent of $Q^2$.
\item The uncertainty on the photon energy measurement is estimated using the BH and DVCS 
control samples.
  For the BH events the cluster transverse energy is compared with the 
track measurement. 
For DVCS events the energy  of the cluster is compared to the energy calculated using the double angle method~\cite{da}. 
The photon energy scale uncertainty estimated 
with this method varies from $1$\% for photons detected in the backward region to $4$\% for forward photons. 
The resulting error on the total inclusive cross section is $+3.6$\%   and $-2.6$\%.
\item An uncertainty of $3$~mrad ($4$~mrad for $\eta_\gamma > 1.4$) is 
attributed to the measurement of the polar angle of the photon. 
This uncertainty is determined by comparing the polar angle measurements of the track and the cluster for identified  electrons in the BH control sample. 
The resulting error on the total inclusive cross section is $+0.1$\%  and $-0.7$\%.
\item  An uncertainty of up to $2$\% is attributed to the energy of the
  scattered electron and an uncertainty of $2$~mrad to the measurement of the scattering angle. The resulting error on the total inclusive cross section is $+1.9$\%  and $-2.9$\%.
\item A $3$\% uncertainty is attributed to the energy of hadronic final
state objects~\cite{highq2}.
 This affects the total cross section  by $+1.2$\%   and $-0.7$\%.
\item A $5$\% uncertainty is applied for the correction of the cluster overlaps in the signal extraction procedure,
   corresponding to half the size of the correction.
\item  The uncertainty attributed to the model dependence of the acceptance corrections is derived
 from the differences between the acceptances determined with the 
 PYTHIA and HERWIG   simulations. A resulting error of  $5$\%
is found  for the inclusive and the \jets  and
$10$\% for the \nojet cross sections. 
\item The uncertainty of the ratio of the \LL and \QQ contributions
for the acceptance corrections
 is taken into account by varying the scaling factor for PYTHIA from $1.5$ to $3$. 
The resulting error on the total inclusive cross section is $+1.4$\%  and $-0.8$\%.
For the double differential cross section, a 
systematic error of up to $5$\% is found. 
In regions with $\eta^\gamma >-0.6$, the systematic error is below $1$\%.
\item An uncertainty of $1$\% is attributed to the simulation of the trigger efficiency.
\item 
The uncertainty on the track  reconstruction  efficiency results in an error of $\pm 0.3$\% for the total inclusive cross section.
\item The uncertainty on the conversion probability of the
        photons before entering the calorimeter results in a systematic error 
of $2$\% of the
         cross sections.
\item The uncertainty on the luminosity measurement is $3.4$\%.
\end{itemize}
In each analysis bin the individual effects of these experimental 
uncertainties are combined in quadrature. 
The  systematic  uncertainty obtained on the total inclusive cross section is 
 $+13.6$\%  and $-15.5$\%.
The largest contribution to this uncertainty arises
from the systematic effects attributed to the description of the shower shapes,
which is partially correlated among measurement bins.
\section{Theoretical Predictions}
The measured cross sections are compared to
the unscaled MC prediction as discussed in section~3. In addition, data are compared with fixed order QCD calculations, described in this section.
 A LO $\mathcal{O}(\alpha^3\alpha_s^0)$ calculation~\cite{Thomas,Thomas1} is used. 
The photon plus jet cross sections are further compared to a NLO $\mathcal{O}(\alpha^3\alpha_s)$ QCD calculation~\cite{kramer} which 
is only available for the photon plus jets phase space selection. 
In the calculations, the  cross section $\sigma(ep\rightarrow e\gamma X)$ is obtained by convoluting the parton-level  cross sections (for instance $ \hat{\sigma} (eq \rightarrow e\gamma q)$ at LO)  with the proton parton density functions. 
\par
The \QQ contribution is dominated by the direct radiation of the photon from the quark involved in the
   parton level process, but also contains  the contribution from  quark fragmentation to a  photon~\cite{Koller:1979df,Gluck:1994iz}.
The direct part can be calculated in perturbation theory. 
The fragmentation contribution is described by a DIS matrix element $eq \rightarrow eq$ convoluted with a process 
independent quark-to-photon fragmentation function determined from data and denoted by $D_{q\rightarrow\gamma}(z)$.
Here $z$ is the  fraction of the quark's momentum carried by the photon. 
Already at leading order, the parton-level cross section contains an infrared divergence 
due to the emission of a photon collinear to the quark.
It is factorised  into the fragmentation function at a factorisation scale $\mu_{F}$.
This singularity at LO makes a  NLO prediction for the inclusive cross section 
considerably involved.
\par
So  far, only two measurements exist that give direct information on the quark-to-photon fragmentation function. 
These measurements were made by the LEP experiments  ALEPH~\cite{Buskulic:1995au} and OPAL~\cite{Ackerstaff:1997nh}.
Only the ALEPH measurement has the precision to allow a determination of $D_{q\rightarrow \gamma}(z)$. 
The LO calculation used in this analysis is based on the ALEPH LO 
parametrisation of   $D_{q\rightarrow \gamma}(z)$~\cite{Glover:1993xc,Buskulic:1995au}. 
 The  NLO calculation uses the fragmentation function from~\cite{Bourhis}.
 In the \mc models the contribution from fragmentation is not included.
\par
For photons plus no-jets the quark-to-photon fragmentation enters already at lowest   order ($\alpha^3\alpha_s^0$)~\cite{Thomas}, 
in contrast to the \jets sample where the fragmentation contribution is of the order $\alpha^3\alpha_s^1$.
Since the contribution from fragmentation is suppressed by the
requirement of the photon being isolated,         
the present analysis has no sensitivity for a determination of
$D_{q\rightarrow \gamma}(z)$.
\par
In the \LL subprocess, the photon is radiated by the lepton. 
In the present analysis, the reconstruction of the photon and the electron  in different parts of the
detector ensures that the two particles  are separated, 
hence the \LL term contains no collinear singularities. 
The interference term ($LQ$), which  differs in sign for $e^+p$ and $e^-p$ scattering, is included in the calculations. 
It contributes less than $3$\% to the cross section~\cite{Thomas1}. 
\par
The calculations of the isolated photon cross section are made for  the same ratio of luminosities for $e^+p~(47.9$\%$)$ and $e^-p~(52.1$\%$)$ scattering as for the data.
The cuts in the theoretical calculations are adjusted to correspond to the 
experimental  cuts described in this paper. The calculations use the same jet algorithm, applied on parton level, as used for the data analysis.
The CTEQ6L~\cite{Stump:2003yu} leading order parametrisation of proton parton distributions is used.
Different proton parton density functions are found to change the predictions by  $5-10$\%.
The LO and NLO predictions are compared to the data after a correction for hadronisation effects.
The correction factors $f_{had}$ are defined as the ratio of the cross sections calculated from 
hadrons to those from partons and are  determined  from the scaled signal MC. 
The same jet algorithm as for the data is applied on parton and on hadron level.
The uncertainty of the correction factor is estimated by comparing the correction factors 
obtained from PYTHIA with those from HERWIG.
The correction for the total inclusive cross section is  $-14$\%  with an associated  uncertainty 
of $5$\%. For the differential cross sections the corrections are at  most $-30$\%. The correction is largest for low photon energies and in forward direction of the photon because of the
degraded resolution of the isolation parameter $z$.
The combined uncertainties of the theoretical predictions from hadronisation corrections and proton 
parton distributions amount to up to $11$\%.

The NLO calculation of the photon plus jet cross sections includes
processes with an additional gluon, either as the incoming parton or in the final state, as well as virtual corrections.
The renormalisation and factorisation scales are set to $\mu_R\!=\mu_F=\!\sqrt{Q^2 +(P_T^{jet})^2} $. 
Theoretical uncertainties are estimated by varying $\mu_R$ and $\mu_F$ independently by a factor two up and down.
 These uncertainties 
 are below $3$\% and lower than the uncertainties 
from the choice of the  proton parton distributions and from the hadronisation corrections.

\section{Results}
%
The isolated photon cross sections presented below are given for the phase space
defined in table~\ref{tab:kinem}.
Bin averaged differential cross sections are presented
in figures~\ref{fig:inclxsecsingle} to~\ref{fig:xsecsingle1jet} and in  tables~\ref{tab:inclusive} to~\ref{tab:jets}.
For all measurements the total uncertainty is dominated by systematics.
The results are compared with the signal MC predictions 
(unscaled PYTHIA plus RAPGAP) and with the LO and NLO calculations.
The factors $f_{had}$ for the correction of the theoretical calculations from parton
to hadron level are given in the cross section tables. 

\subsection{Inclusive isolated photon cross sections}
The measured inclusive isolated photon cross section  is
\begin{displaymath}
\sigma(ep\rightarrow e \gamma X) = 50.3 \pm 1.7 \;(\mbox{stat})\;^{+6.8}_{-7.8} \;\mbox{(syst)}\ \mbox{pb.}
\end{displaymath}
The LO calculation predicts a cross section of $28.6$~pb, while the
  \SumMC expectation is $26.4$~pb.
Thus both predictions significantly underestimate the measured total inclusive cross section
by almost a factor of two.
\par
\begin{figure}[hhh]
    \includegraphics[width=0.473\textwidth]{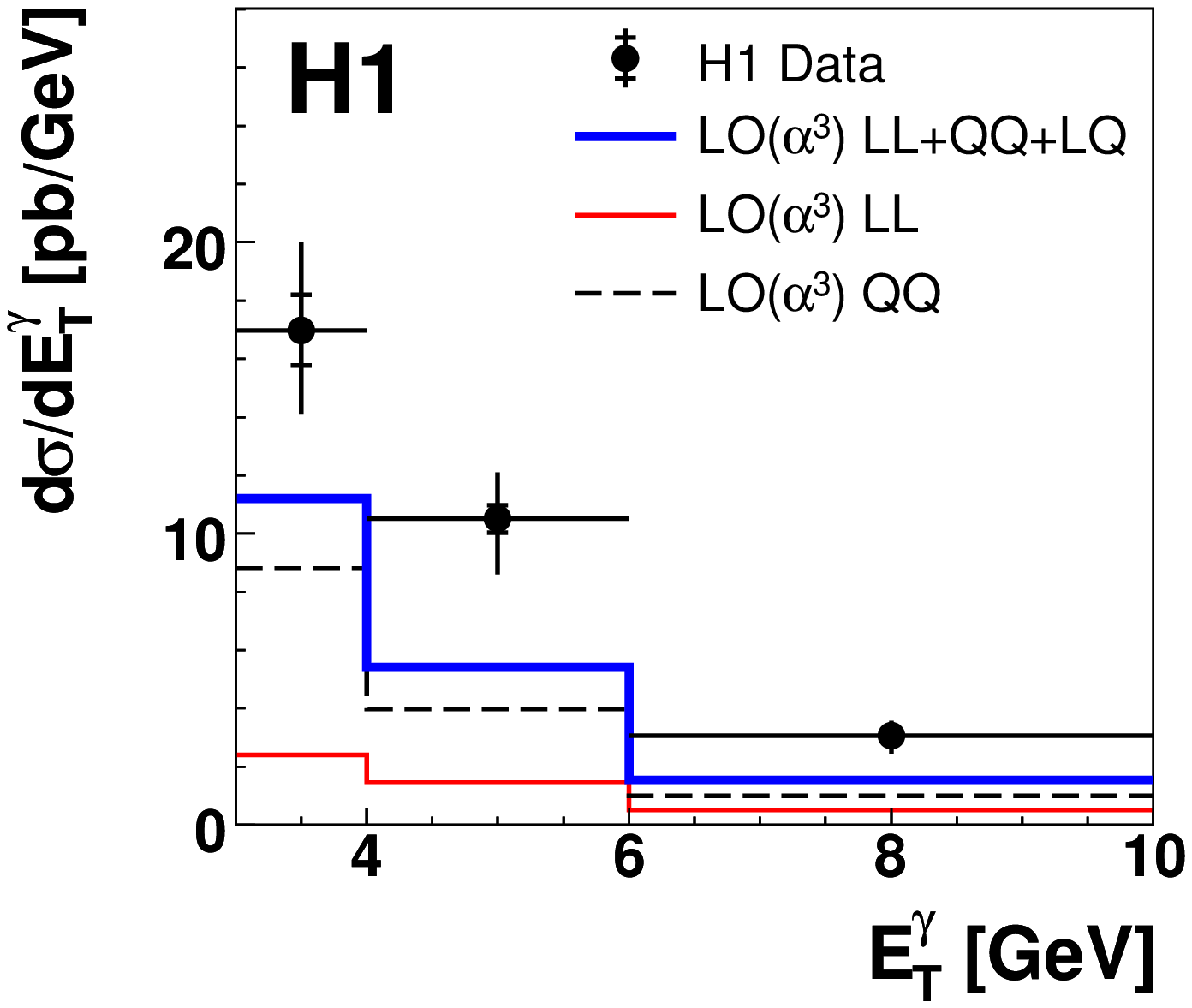}
    \vspace{0.3cm}
    \hspace{0.04\textwidth}
    \includegraphics[width=0.473\textwidth]{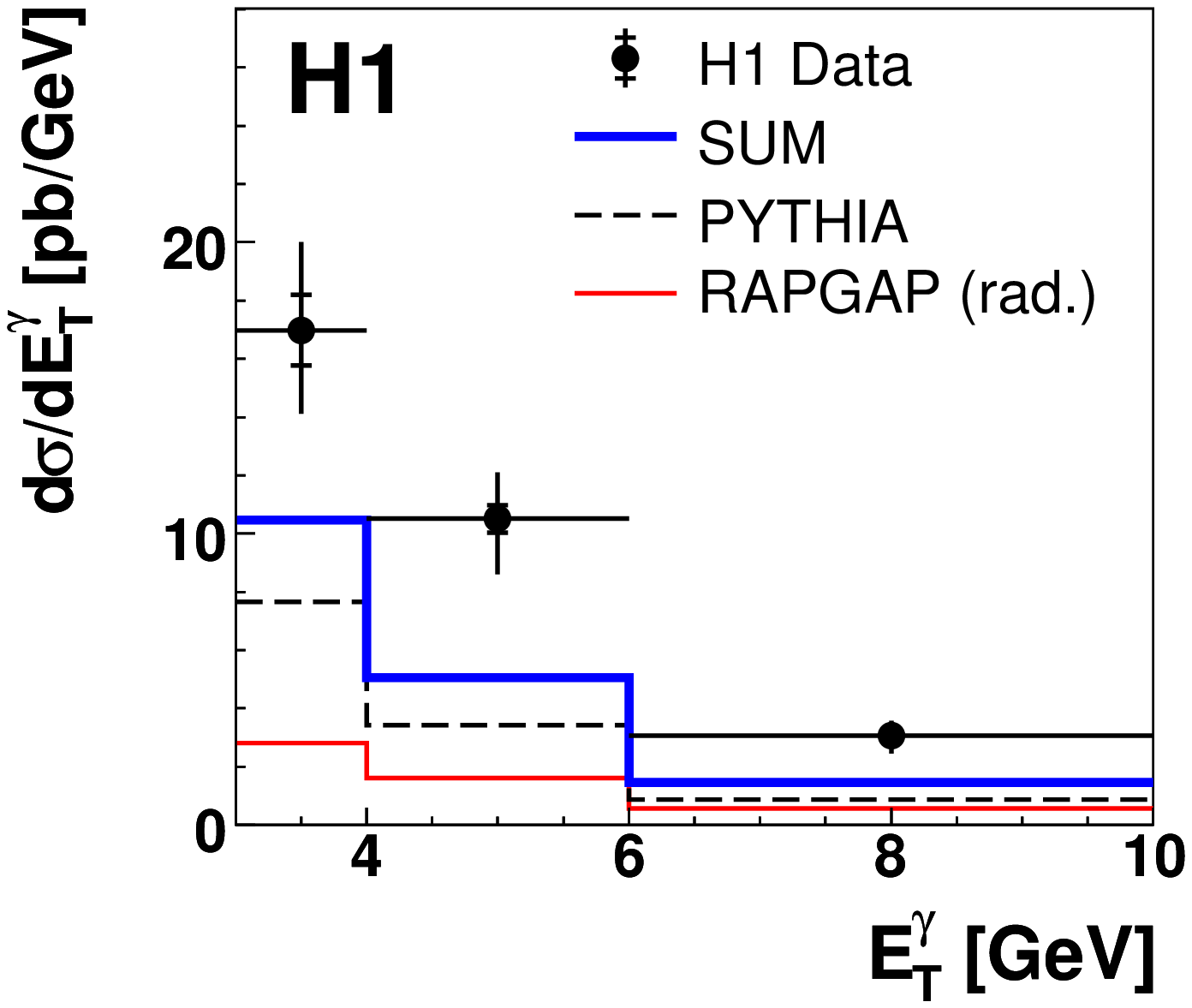}
    \includegraphics[width=0.473\textwidth]{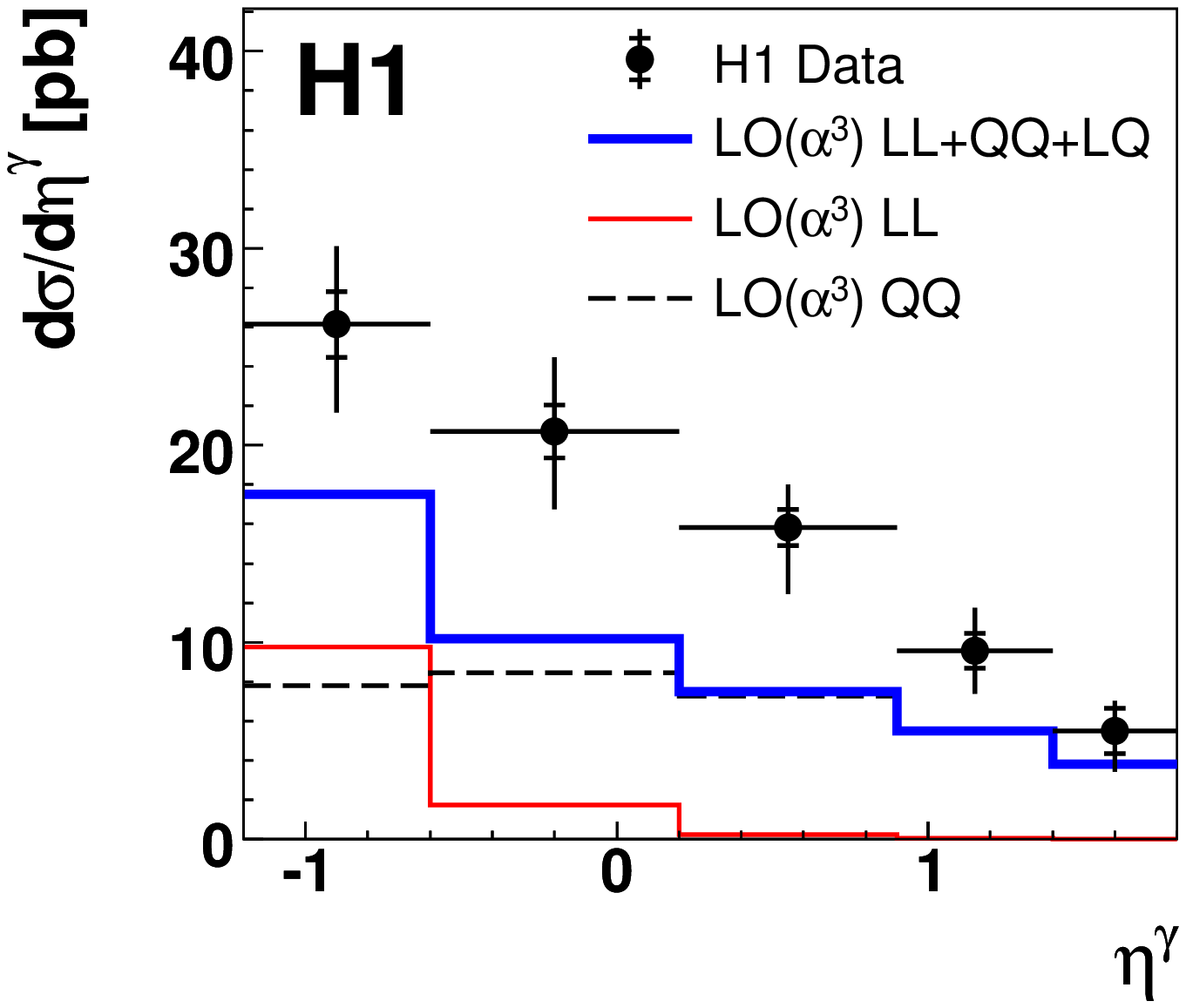}
    \vspace{0.3cm}
    \hspace{0.04\textwidth}
    \includegraphics[width=0.473\textwidth]{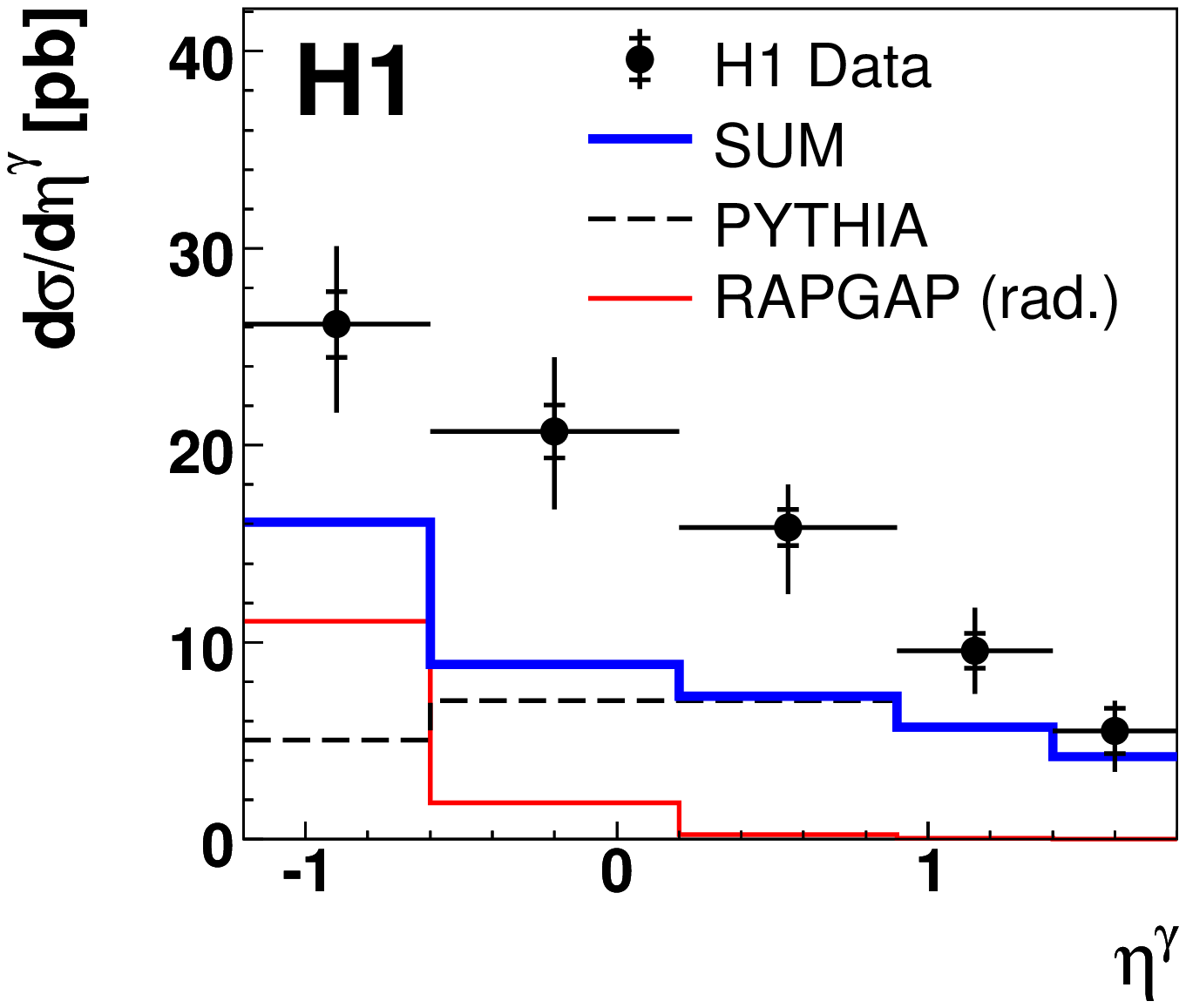}
    \includegraphics[width=0.473\textwidth]{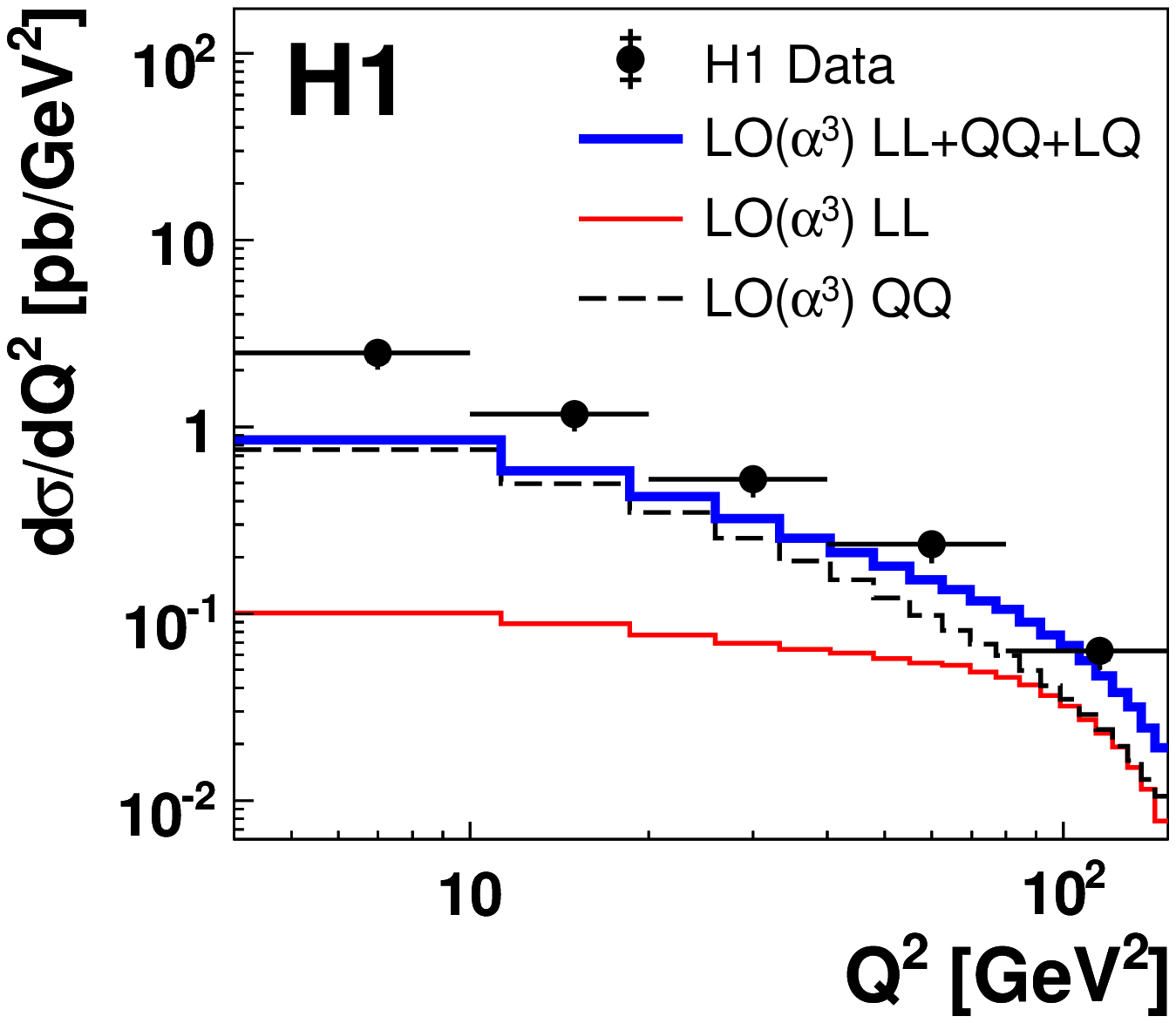}
    \hspace{0.04\textwidth}
    \includegraphics[width=0.473\textwidth]{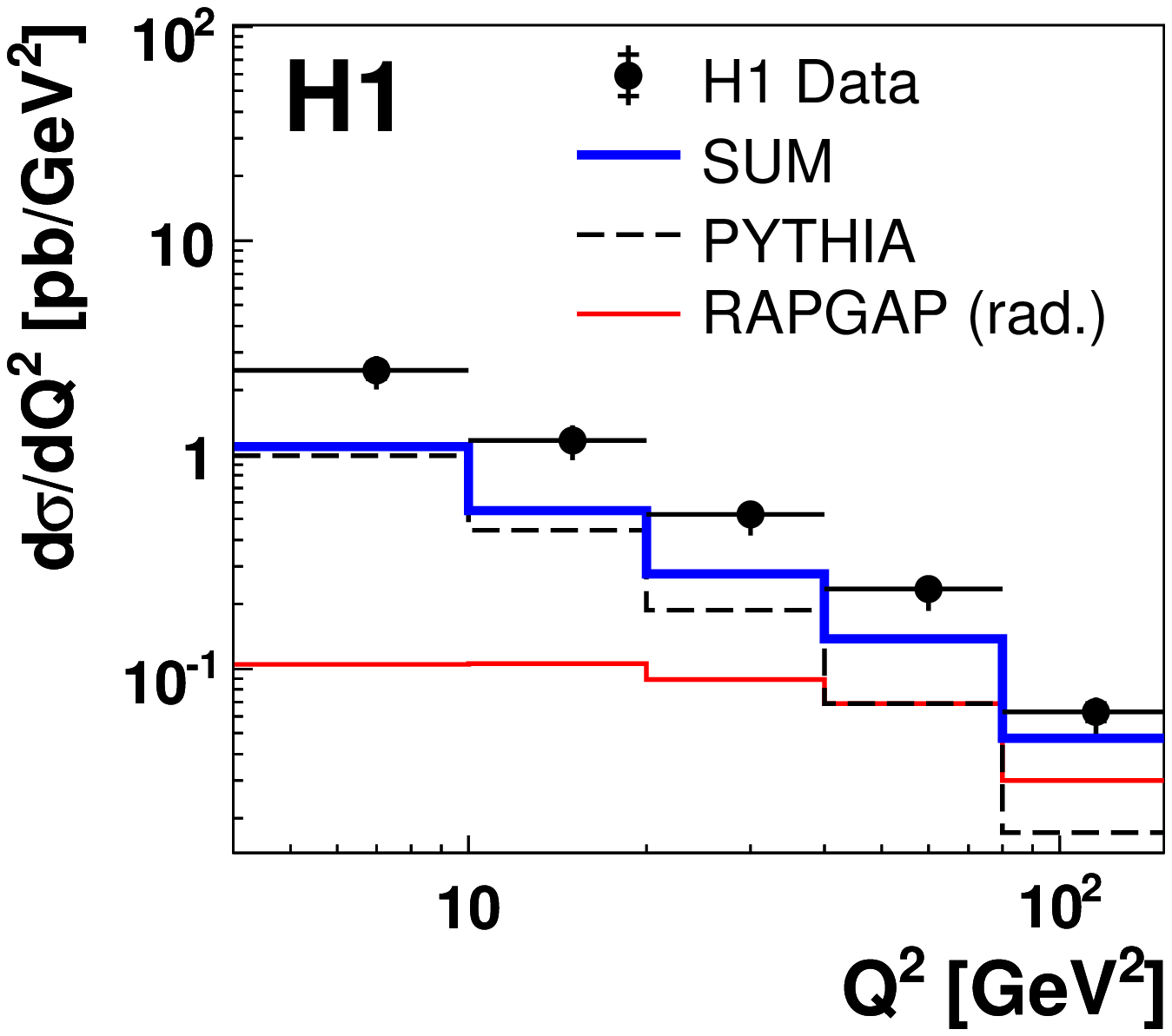}
  \begin{center}
    \vspace*{-0.2cm}
    \caption{Differential cross sections for inclusive isolated photon production 
     $d\sigma/dE_T^{\gamma}$, $d\sigma/d\eta^{\gamma}$,
     and $d\sigma/dQ^2$ 
in the kinematic range specified in table~\ref{tab:kinem}.
The inner error bars on the data points indicate the statistical error,
 the full error bars contain in addition the systematical errors added in quadrature.
The cross sections in (a, c, e) are shown together with a leading
order, $\mathcal{O}(\alpha^3\alpha_s^0)$, calculation corrected for hadronisation effects, 
\LL corresponding to radiation from the electron, \QQ to radiation from
the quark and \LQ to their interference. The same cross sections are shown in (b, d, f)
together with the prediction from PYTHIA for photon emission from the
quark  and from RAPGAP~(rad.) for emission from the electron.}
    \label{fig:inclxsecsingle}
  \end{center}
  \begin{picture}(0,0)
    \put(42,258){\textsf{{\large H1 Isolated Photon Production in DIS}}}
    \put(66,246.9){\textsf{(a)}}
    \put(150.5,246.9){\textsf{(b)}}
    \put(66,181.9){\textsf{(c)}}
    \put(150.5,181.9){\textsf{(d)}}
    \put(66,116.4){\textsf{(e)}}
  \put(150.5,116.4){\textsf{(f)}}
  \end{picture}
\end{figure}

Differential cross sections $d\sigma/dE_T^\gamma$, $d\sigma/d\eta^\gamma$ and
$d\sigma/dQ^2$ are presented in table~\ref{tab:inclusive} and in figure~\ref{fig:inclxsecsingle}.
\par

The data are compared in the left panels of figure~\ref{fig:inclxsecsingle}
with the LO predictions, displaying
separately the \LL and \QQ contributions.
%
%
The LO calculation provides a reasonable description of the shapes of the data distributions in
$E_T^\gamma$ and $\eta^\gamma$, while the global normalisation is about a factor of two too low.
%
%
The calculation is closest to the data at larger $Q^2$ and for backward photons
($\eta^\gamma<-0.6$), where the \LL  contribution 
is of similar magnitude to that of $QQ$. For forward and central photons ($\eta^\gamma>-0.6$), the \QQ contribution dominates.
%
%
%
\par
The measurements are compared in the right panels of
figure~\ref{fig:inclxsecsingle}
with the signal MC predictions.
The estimations of the \QQ processes by PYTHIA and of the \LL contributions
by RAPGAP agree well with the predictions from the LO calculation.
Thus the conclusions for the comparison of the signal MC with the data
are similar as for the LO calculation.
This agreement between LO and the signal MC
holds for all cross sections studied in this analysis
and in the following the data are only compared with the 
LO calculation.
%
\par
\begin{figure}[hhh]
    \includegraphics[width=0.473\textwidth]{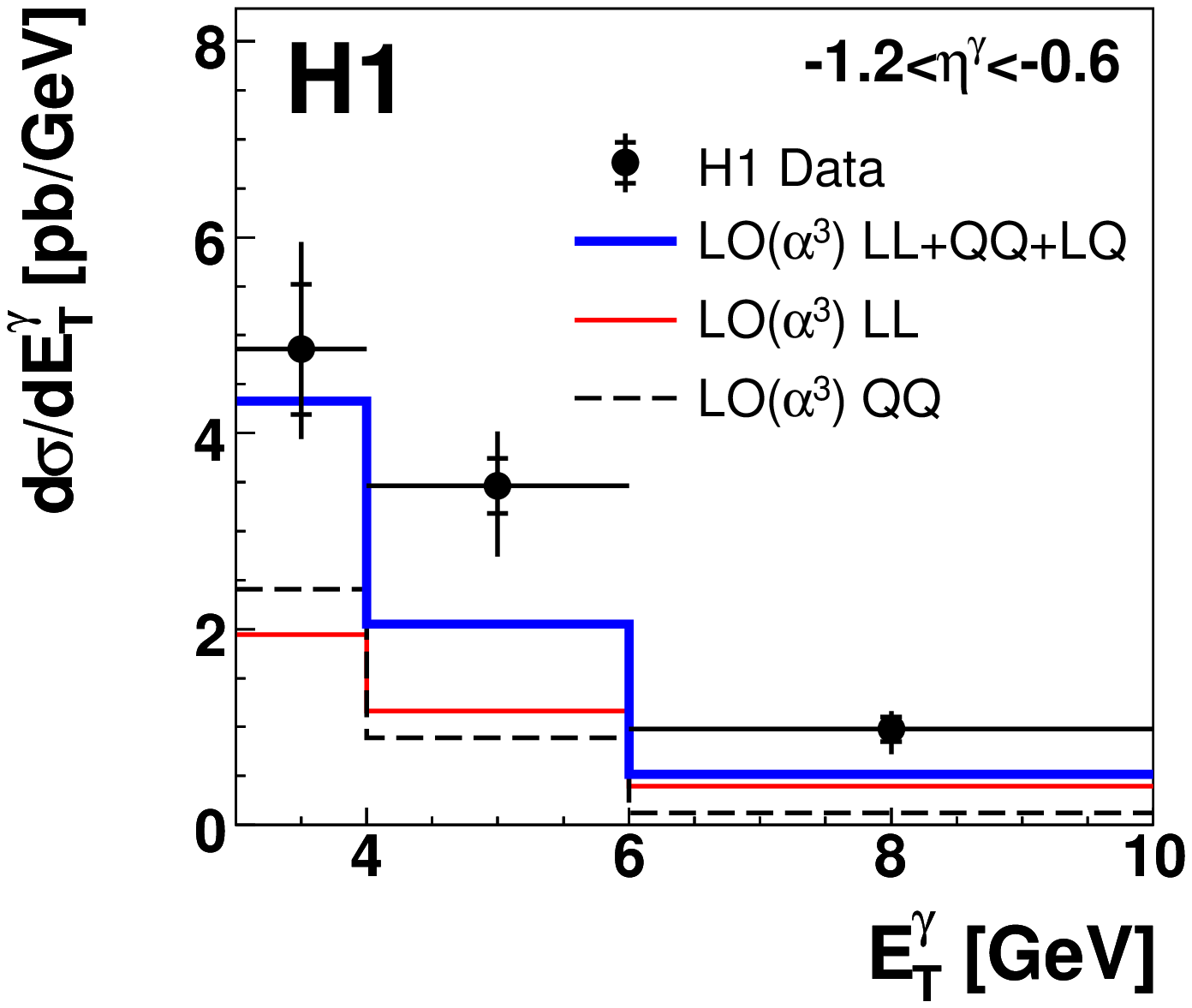}
    \vspace{0.3cm}
    \hspace{0.04\textwidth}
    \includegraphics[width=0.473\textwidth]{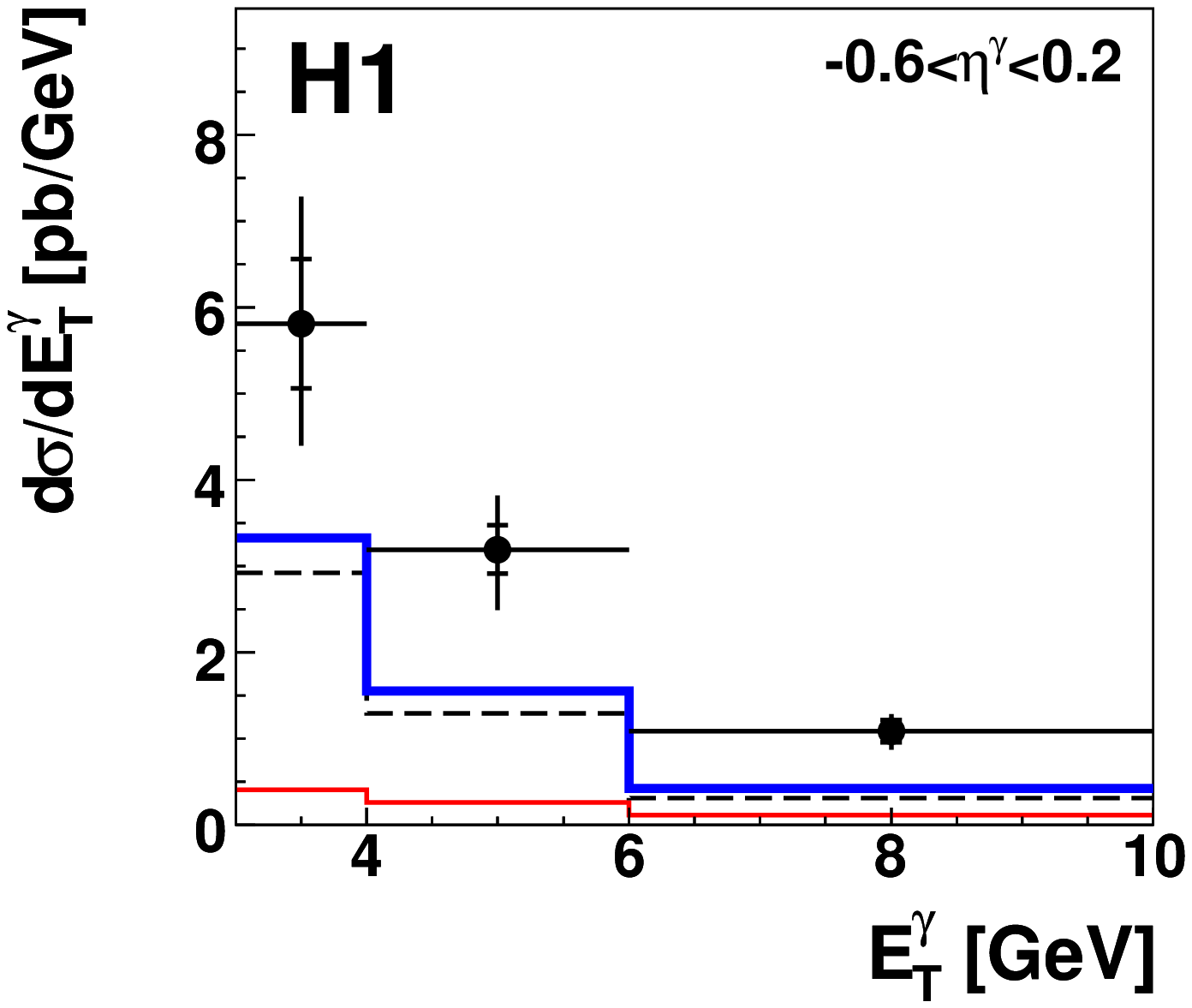}
    \includegraphics[width=0.473\textwidth]{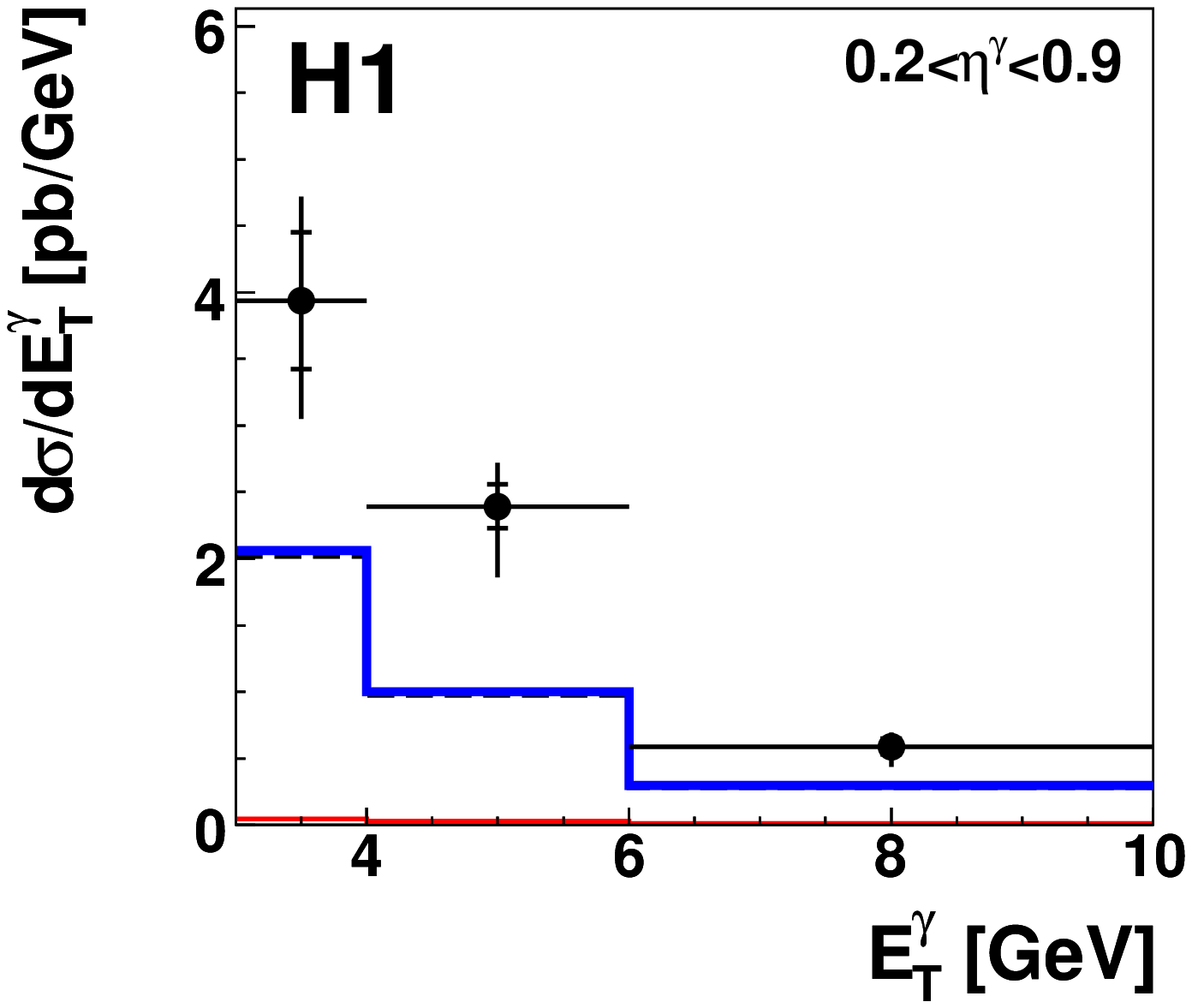}
    \vspace{0.3cm}
    \hspace{0.04\textwidth}
    \includegraphics[width=0.473\textwidth]{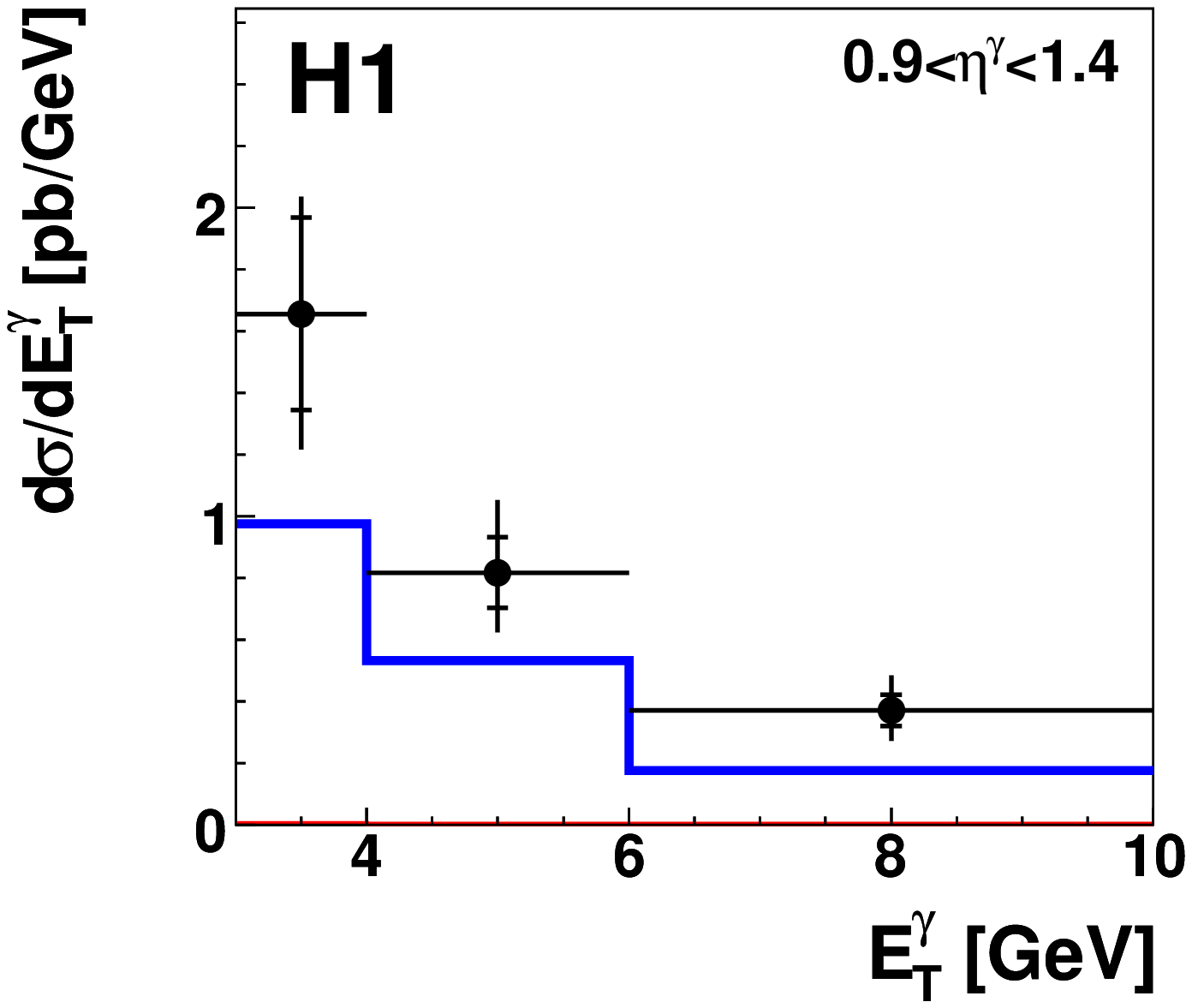}
    \includegraphics[width=0.473\textwidth]{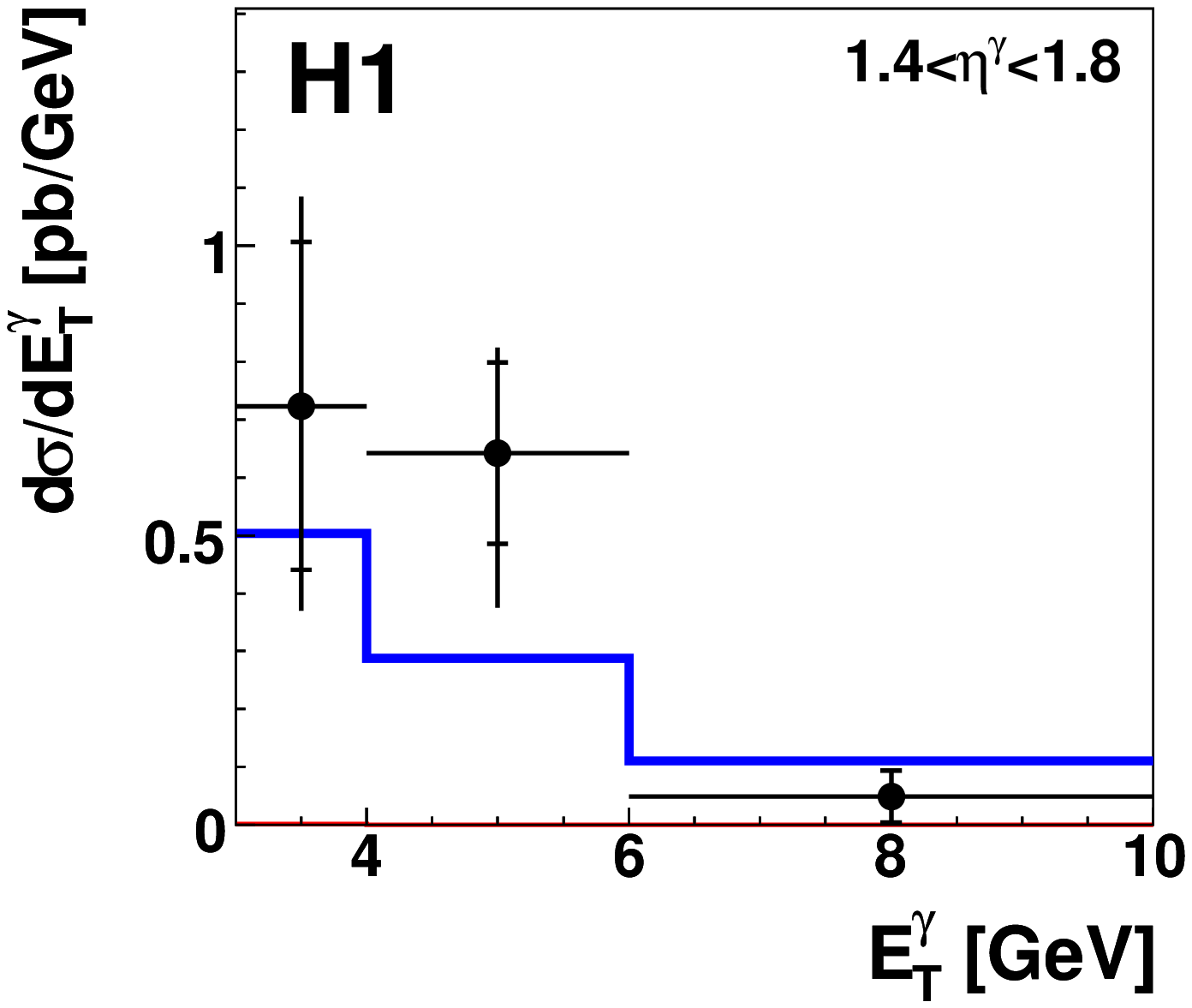}
  \begin{center}
    \caption{Differential cross sections  $d\sigma/dE_T^{\gamma}$ for inclusive isolated photon production
in the kinematic range specified in table~\ref{tab:kinem},
 in 
    $\eta^{\gamma}$~bins corresponding to the wheel structure of the LAr
    calorimeter (see text). 
The measurements are  compared to a leading order $\mathcal{O}(\alpha^3\alpha_s^0)$ calculation (more details
 in the caption of figure~\ref{fig:inclxsecsingle}). The \LL contribution is negligible for $\eta^\gamma>0.2$. } 
    \label{fig:inclxsecdouble}
  \end{center}
  \begin{picture}(0,0)
    \put(42,240){\textsf{{\large H1 Isolated Photon Production in DIS}}}
    \put(66,224){\textsf{(a)}}
    \put(150.5,222){\textsf{(b)}}
    \put(66,156.5){\textsf{(c)}}
    \put(150.5,156.5){\textsf{(d)}}
    \put(66,91){\textsf{(e)}}
  \end{picture}
\end{figure}

Figure~\ref{fig:inclxsecdouble} and table~\ref{tab:double} present the differential cross
section as a function of $E_T^\gamma$ in five different bins of $\eta^\gamma$, corresponding to the wheel structure of the LAr calorimeter. The shapes of all cross section distributions are reasonably well described
by the LO calculation.
A good description of the data can be obtained by a global scaling
of the \QQ contribution by a factor $2.3$ and leaving the \LL normalisation
unchanged.
This is an indication that the observed excess in the data is mainly due
to an underestimation of the \QQ component.
\par
\begin{figure}[hhh]
    \includegraphics[width=0.473\textwidth]{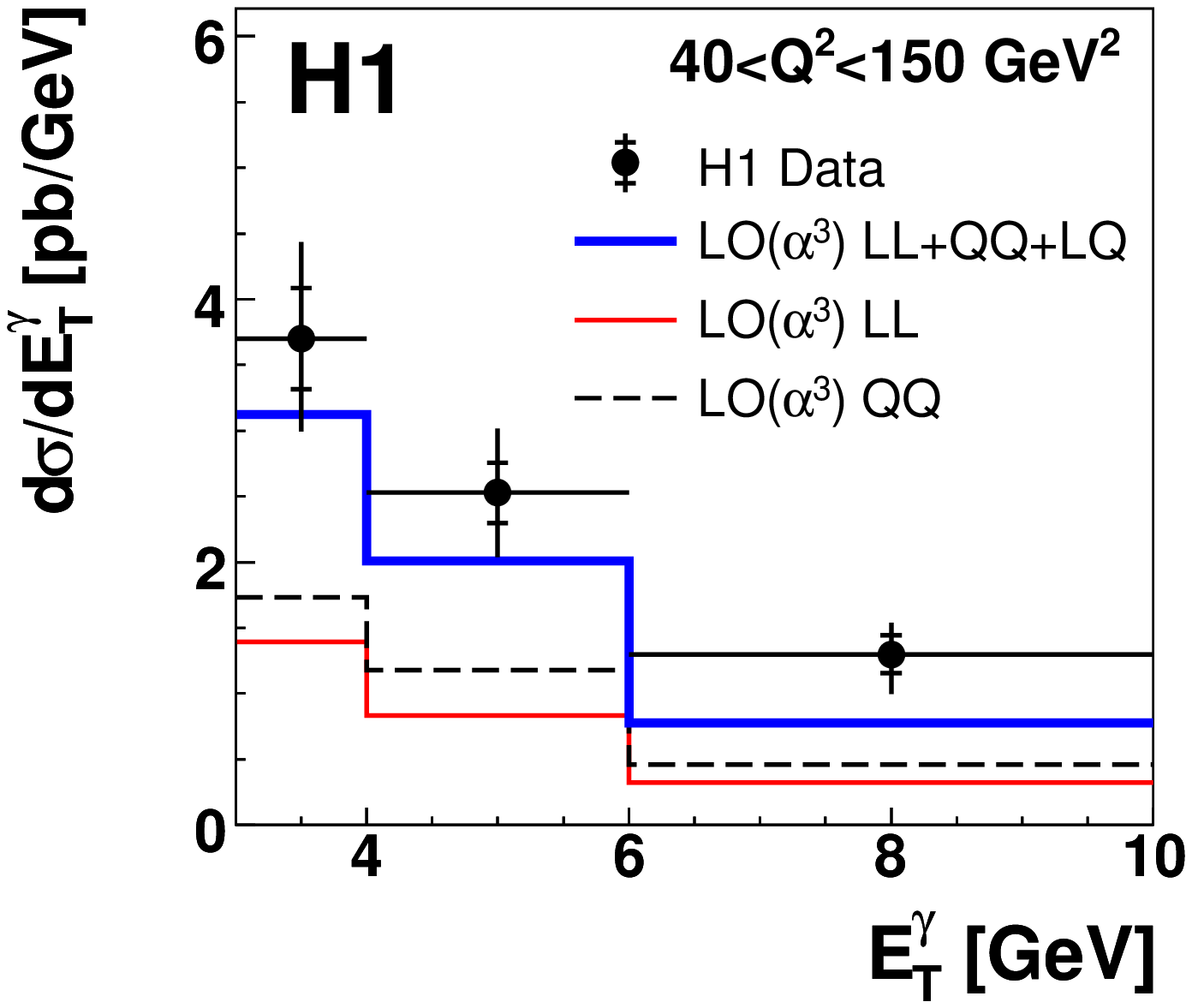}
    \hspace{0.04\textwidth}
    \includegraphics[width=0.473\textwidth]{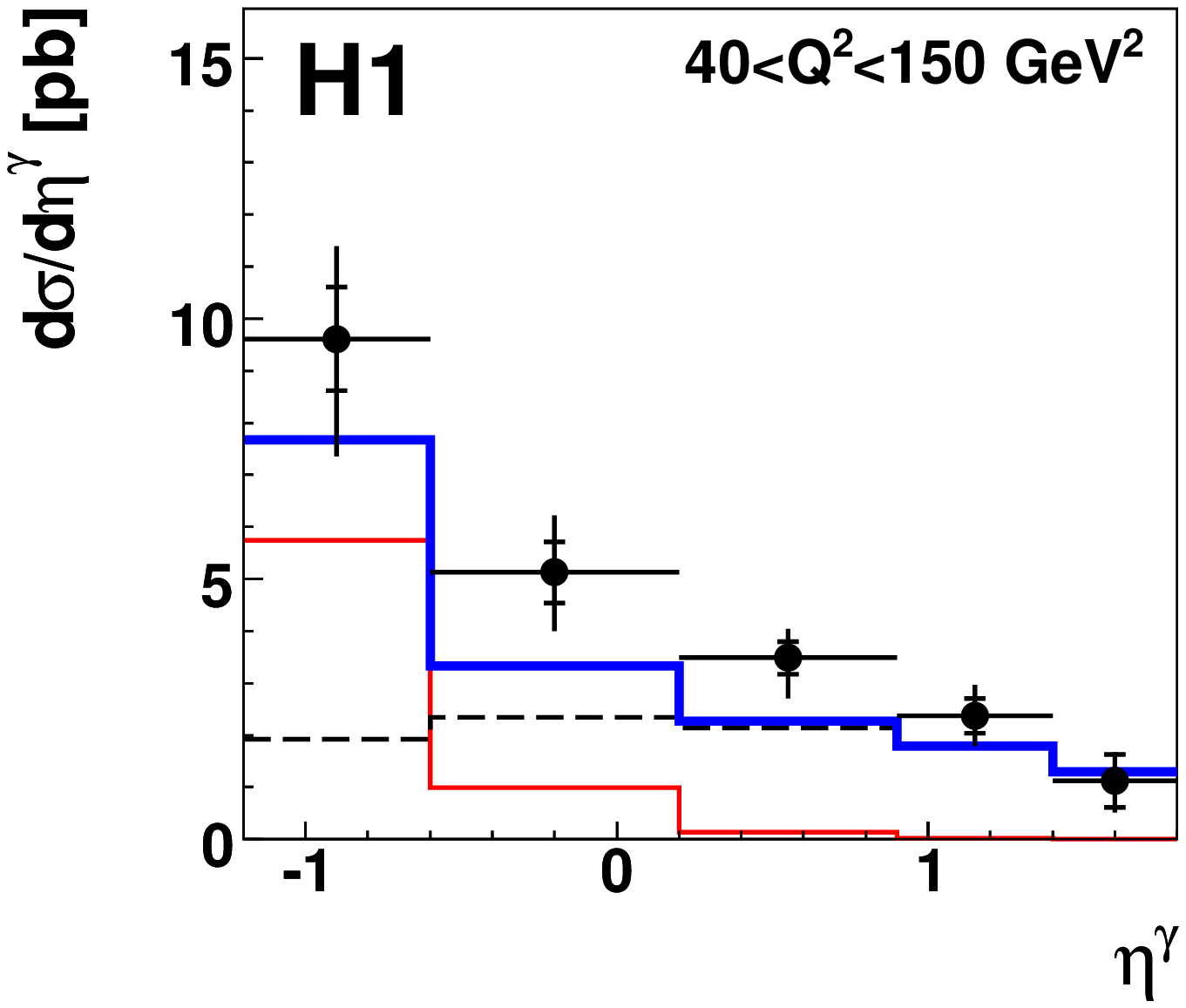}
  \begin{center}
    \caption{Differential inclusive cross sections  $d\sigma/dE_T^{\gamma}$ (a) and
      $d\sigma/d\eta^{\gamma}$ (b) for isolated photon production 
in the kinematic range specified in table~\ref{tab:kinem}  and the additional criterion $40<Q^2<150$~GeV$^2$.
 The cross sections are compared to the leading order, $\mathcal{O}(\alpha^3\alpha_s^0)$,
calculation (more details in the caption of  figure~\ref{fig:inclxsecsingle}).}
    \label{fig:xsechighq2}
  \end{center}
  \begin{picture}(0,0)
    \put(25,106.2){\textsf{\large H1 Isolated Photon Production in DIS at
        $\mathsf{Q^2>40}$ GeV$\mathsf{\mbox{}^2}$}}
  \put(65,93.3){\textsf{(a)}}
  \put(149.5,93.3){\textsf{(b)}}
  \end{picture}
\end{figure}

At $Q^2 > 40\,$GeV$^2$, the agreement of the predictions with the data
is somewhat better, as can be seen in  figure~\ref{fig:xsechighq2} and table~\ref{tab:highQ2}.
The total inclusive cross section for  $Q^2 > 40$~GeV$^2$ is 
 \begin{displaymath}
\sigma(ep\rightarrow e \gamma X) = 14.0 \pm 0.8\; (\mbox{stat})\;^{+2.2}_{-2.1}\;\mbox{(syst)}\ \mbox{pb.}
\end{displaymath}
The LO  prediction of $10.3$~pb is about $30$\% below the data.
The shapes of the $d\sigma/d E_T^\gamma$ and   $d\sigma/d \eta^\gamma$  distributions are well reproduced.
The relative contribution of \LL is predicted to be higher than at low  $Q^2$.
\par
\begin{figure}[hhh]
  \begin{center}
    \includegraphics[width=0.55\textwidth]{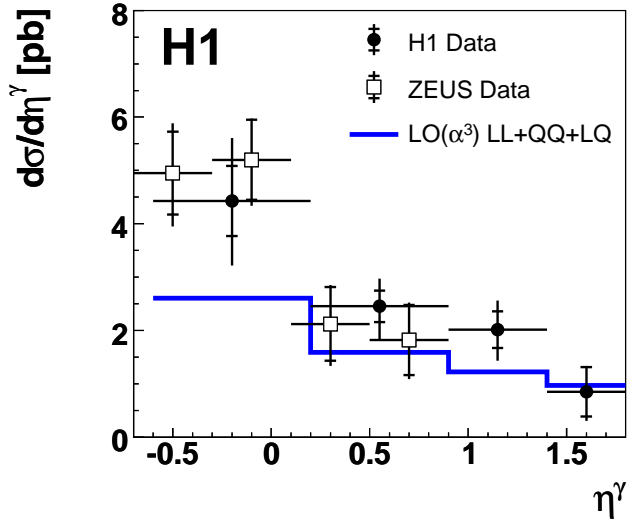}
    \caption{Differential cross sections  $d\sigma/d\eta^{\gamma}$
for the inclusive isolated photon production in comparison to the previous measurement by ZEUS\cite{ZEUSdis} for
$Q^{2}> 35$ GeV$^2$, $E_e'>10$ GeV, $139.8 < \theta_e < 171.9^{\circ}$ and  $5<E_{T}^{\gamma}<10$~GeV. The additional condition  $W_{X}>50$ GeV is used in the H1 analysis only (see section 8).
 The cross sections are compared to the leading order, $\mathcal{O}(\alpha^3\alpha_s^0)$, calculation.
In contrast to the comparison in \cite{Thomas1},  the calculation is here corrected for hadronisation effects.
}
    \label{fig:zeus}
  \end{center}
\end{figure}

The present measurement is extrapolated to the phase space of the 
analysis performed by the ZEUS collaboration~\cite{ZEUSdis}
($Q^{2} > 35$  GeV$^2$, $y>0$, $E_e>10$ GeV, $139.8 < \theta_e < 171.9^{\circ}$ and  $5<E_{T}^{\gamma}<10$~GeV).
Due to  the 
different angular coverage of the calorimeters of the two detectors, the extrapolation is not possible for $\eta^\gamma<-0.6$ because the photon-electron separation cannot be properly taken into account.
Figure~\ref{fig:zeus} shows the comparison of the H1 and ZEUS mesurements of the differential cross section $d\sigma /d\eta^{\gamma}$.
 A good agreement is observed. 
The LO calculation is also shown and is here corrected for hadronisation effects in contrast to the comparison in \cite{Thomas1}.
In the ZEUS analysis the photon radiation from the electron is neglected
in the acceptance corrections and 
no $W_X$ cut is applied.
 A rough estimate shows that with the $W_{X}$ cut used by H1
 and the acceptance calculation with the combination of PYTHIA and RAPGAP,
 the ZEUS cross section values are expected to be  lowered by about $10-30$\%.
A different photon isolation criterion is used in the ZEUS analysis which is not corrected for. 
Studies of isolated photons in photoproduction indicate that the two different 
isolation criteria give very similar results.
The measurement presented in this paper significantly  extends the kinematical region probed by ZEUS in  photon transverse energy and
 pseudorapidity, and in $Q^2$. 

\subsection{Photon plus no-jets and photon plus jet cross sections} 

The cross section for jet production in events with isolated photons
is studied.
The measurement is performed in the phase 
space defined for the inclusive cross section with an additional jet requirement or veto as shown in 
table~\ref{tab:kinem}.
\par
The measured total cross section for \nojet is 
\begin{displaymath}
\sigma(ep\rightarrow e \gamma Y) = 18.8 \pm 1.2\;(\mbox{stat})\;^{+3.3}_{-3.4}\;\mbox{(syst)}\ \mbox{pb,}
\end{displaymath}
\noindent where $Y$ contains no identified hadronic jet. 
The prediction of the LO calculation is $11.7$~pb.
The measured total cross section for the photon plus at least one jet is
\begin{displaymath}
\sigma(ep\rightarrow e \gamma\: jet\: X) = 31.6 \pm 1.2 \;(\mbox{stat})\;^{+4.2}_{-4.8}\mbox{(syst)}\ \;\mbox{pb.}
\end{displaymath}
\par
The prediction of the LO calculation is $16.7$~pb. 
For both samples the predictions are significantly lower than the data.
The observed ratios of data to LO prediction are very similar to the inclusive isolated photon measurement. 
As for the inclusive sample, similar conclusions are found for the MC predictions.

A comparison to a  NLO calculation is possible for the \jets cross sections.
The NLO calculation predicts a cross section of $20.2\pm 0.6$~pb, 
about $20$\% higher than the LO prediction but still roughly $35$\% lower than the data.
The analysis performed using a higher cut on $P_T^{jet}>4$~GeV lead to a similar disagreement between the data and the calculation.

\begin{figure}[hhh]
    \includegraphics[width=0.45\textwidth]{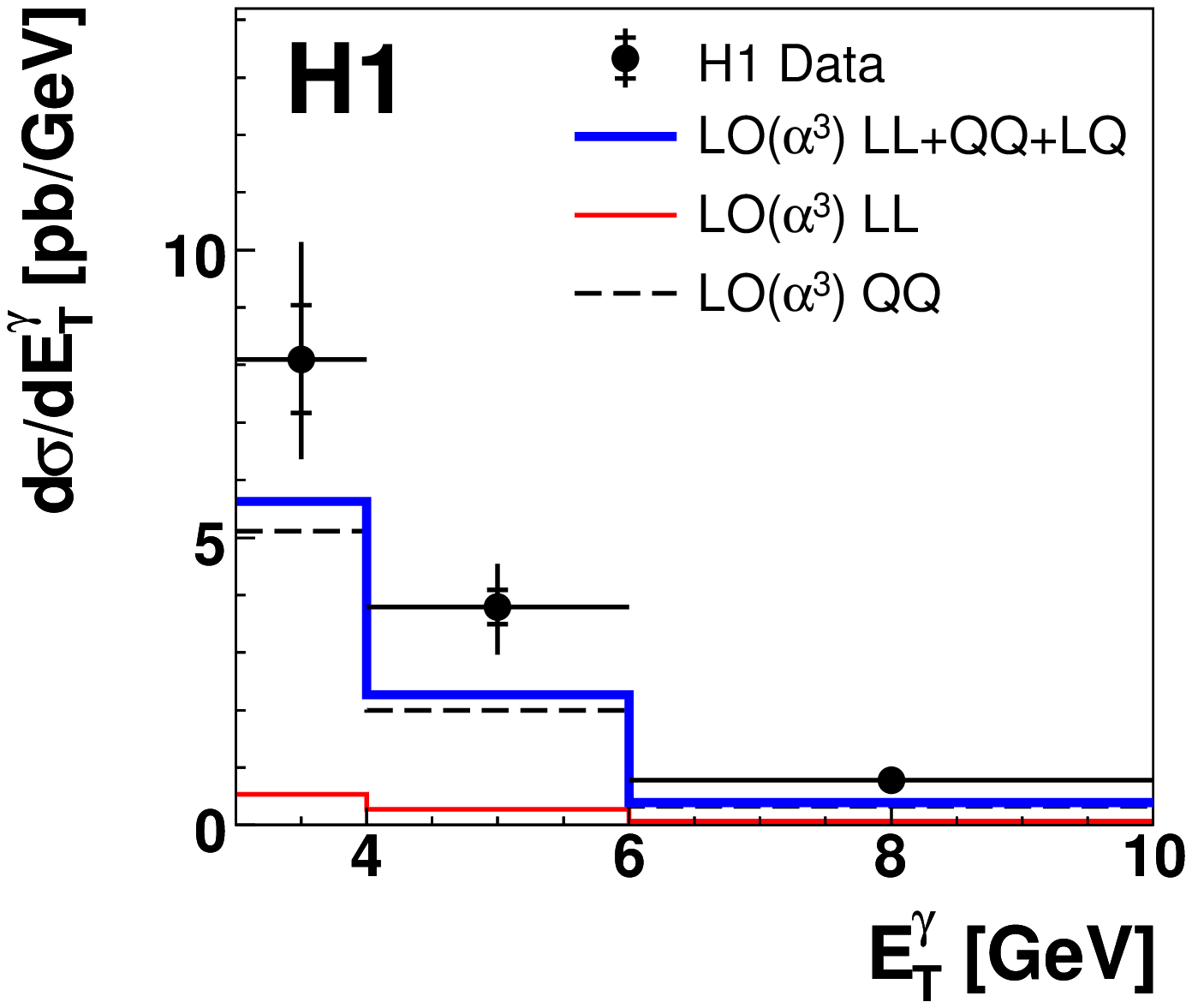}
    \vspace{0.25cm}
    \hspace{0.04\textwidth}
    \includegraphics[width=0.45\textwidth]{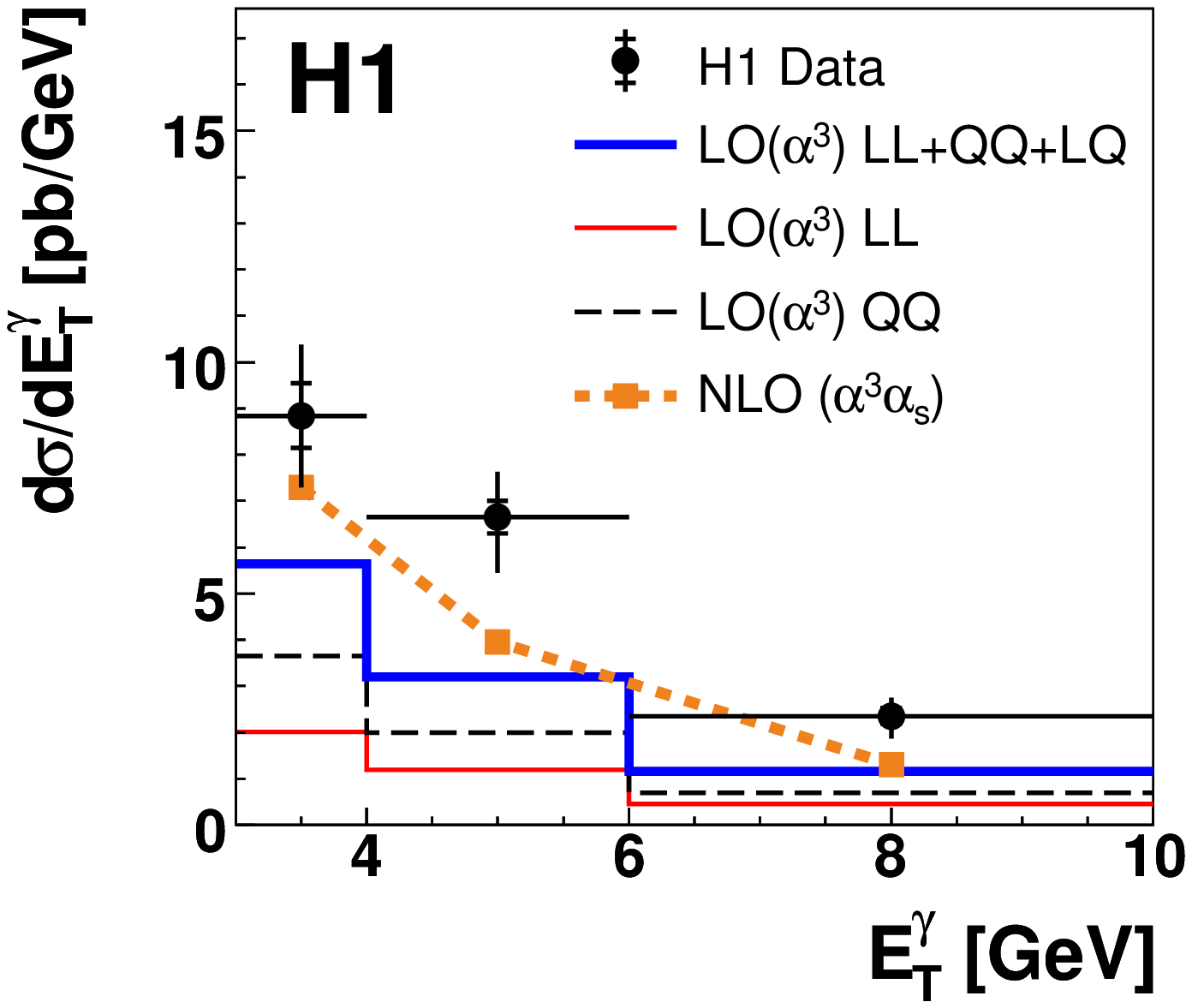}
    \includegraphics[width=0.45\textwidth]{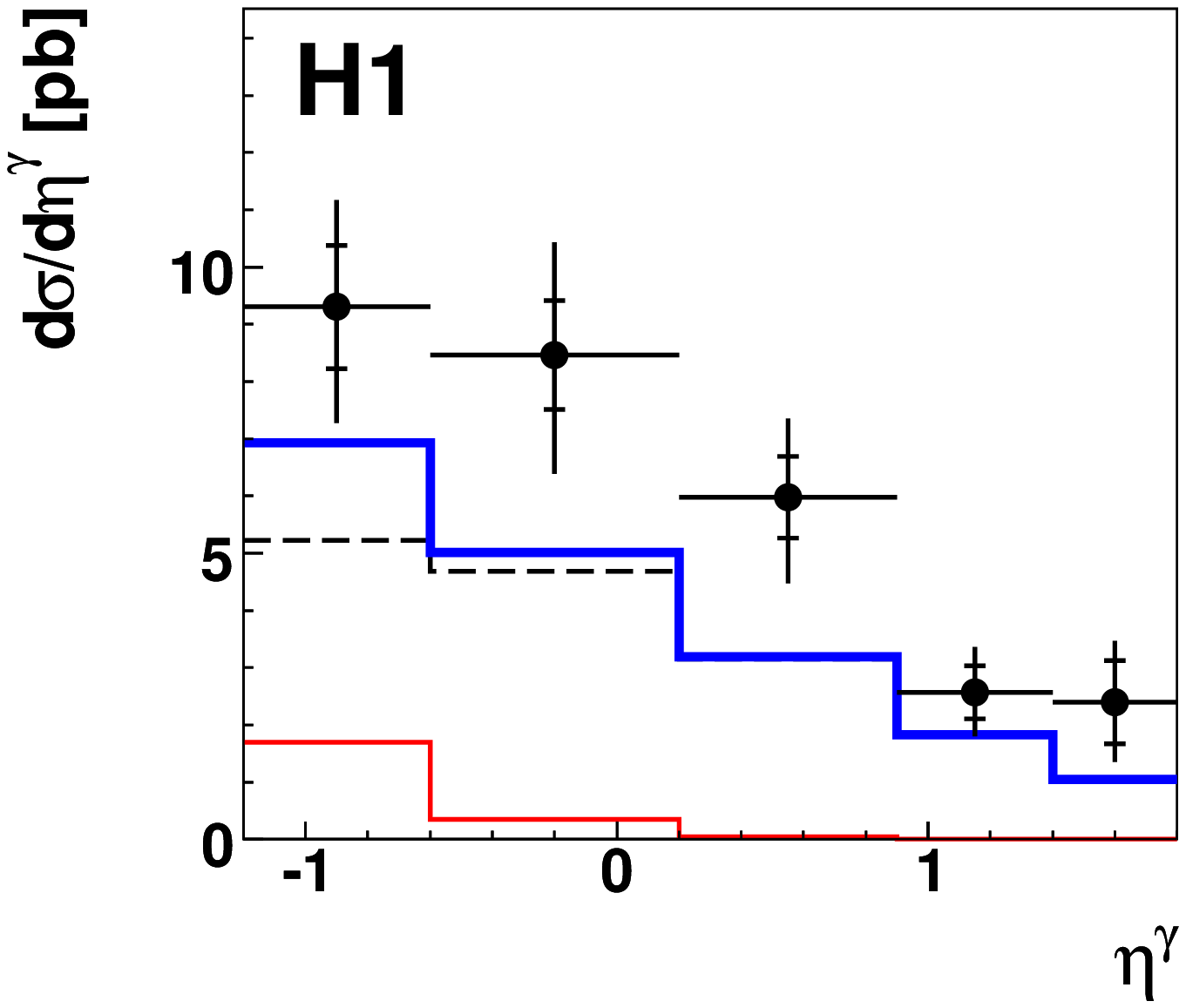}
    \vspace{0.25cm}
    \hspace{0.04\textwidth}
    \includegraphics[width=0.45\textwidth]{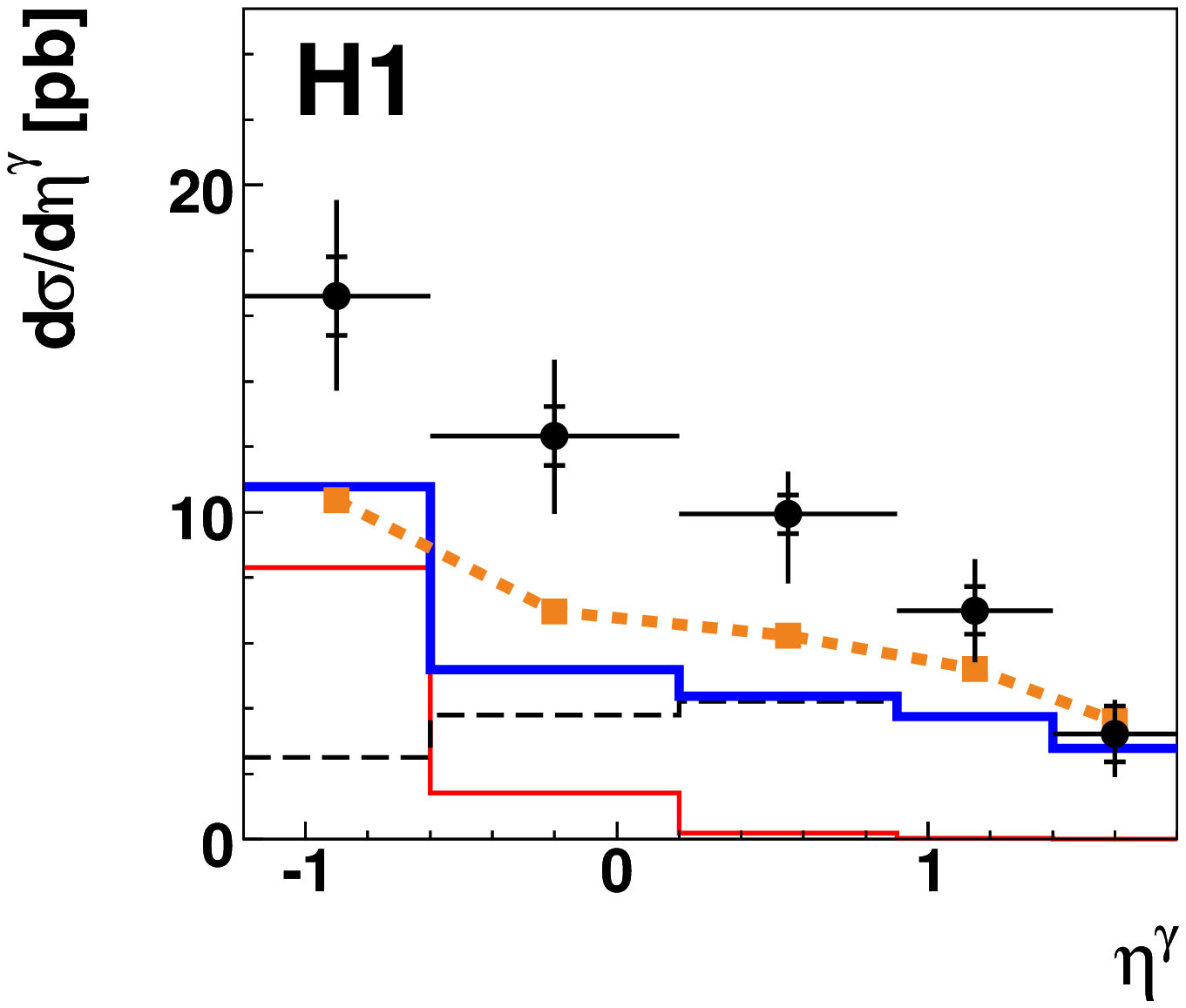}
    \includegraphics[width=0.45\textwidth]{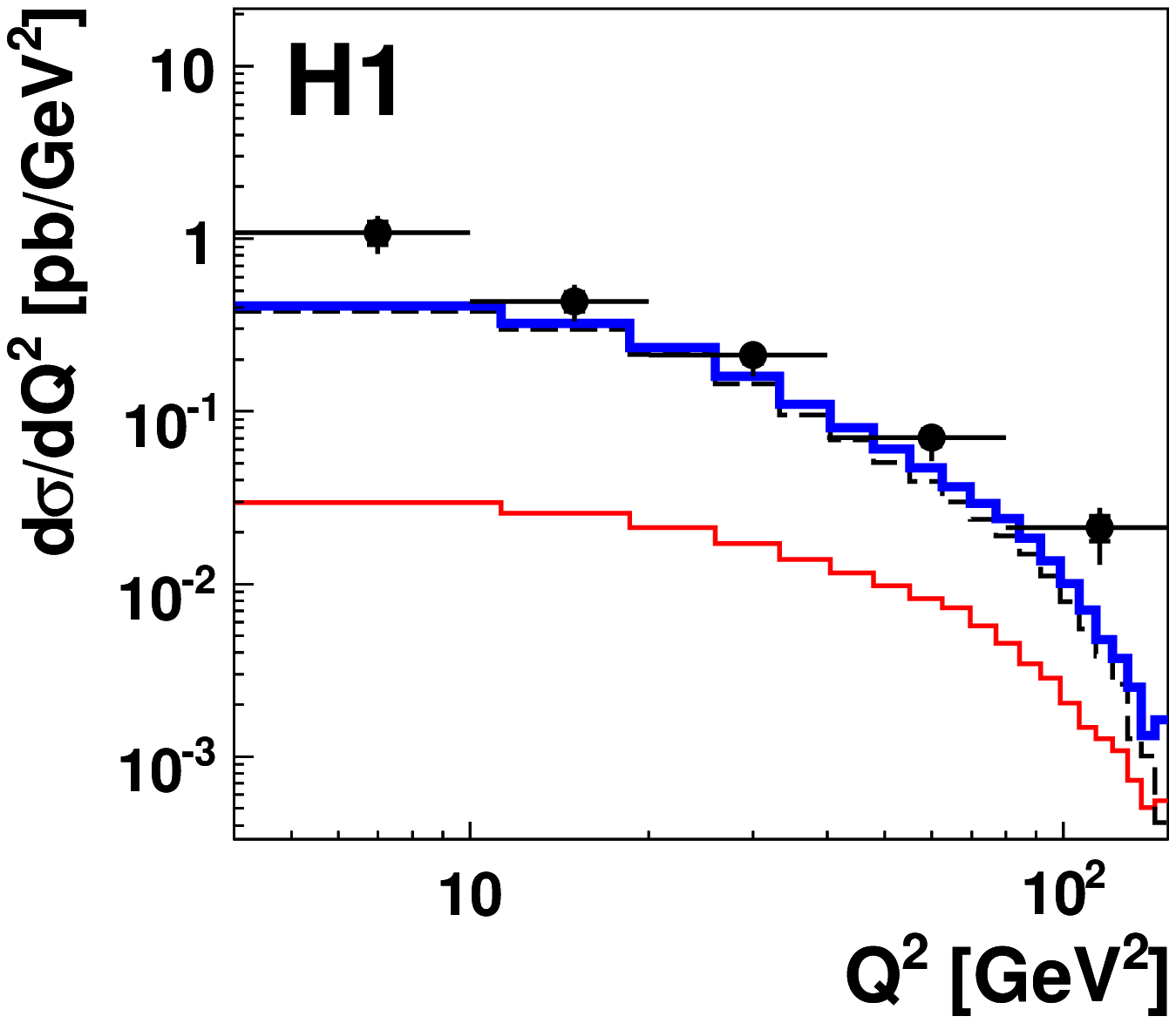}
    \hspace{0.04\textwidth}
    \hspace{0.04\textwidth}
    \includegraphics[width=0.45\textwidth]{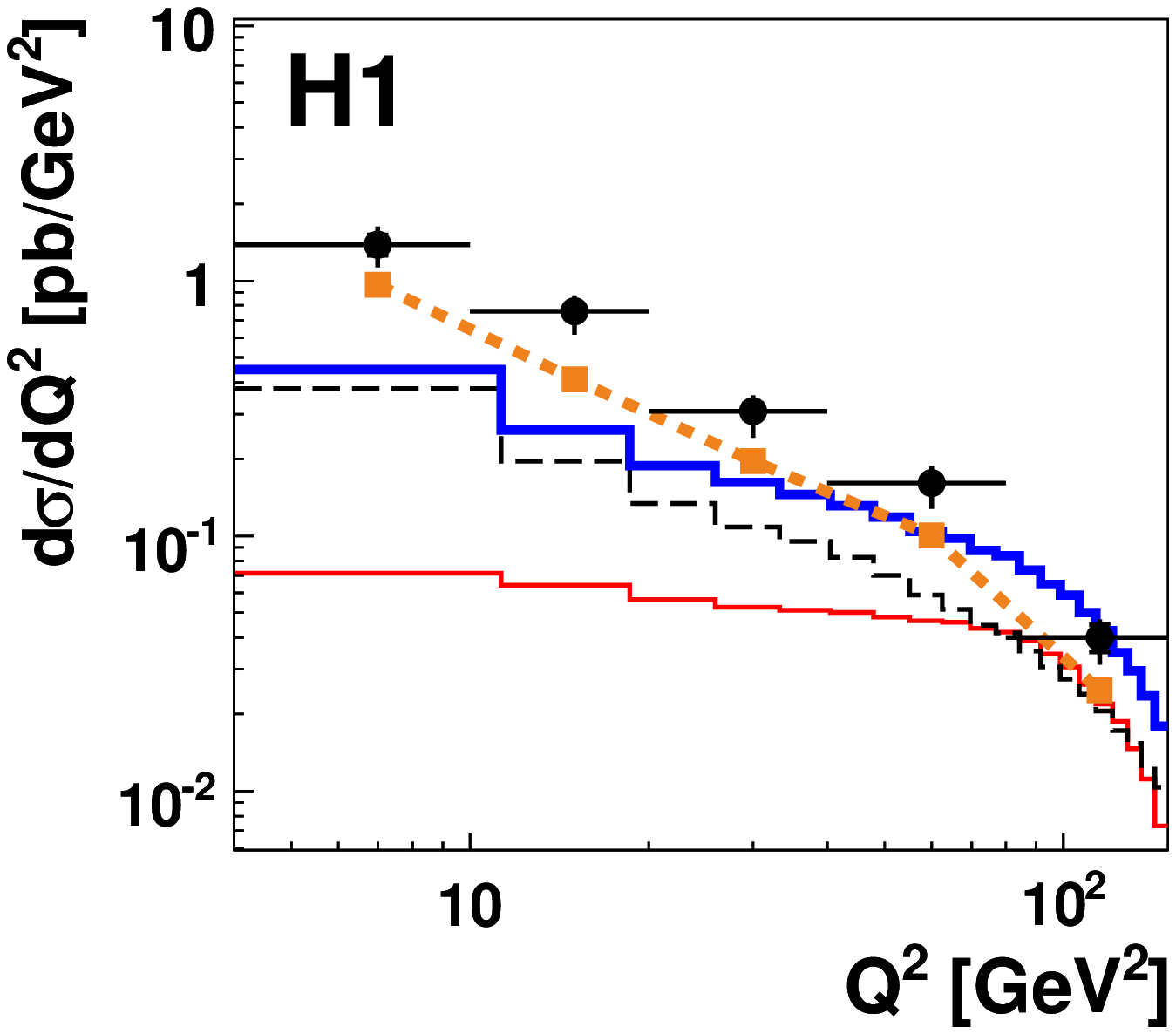}
  \begin{center}
    \caption{Differential cross sections  $d\sigma/dE_T^{\gamma}$, $d\sigma/d\eta^{\gamma}$ and
    $d\sigma/dQ^2$  for  \nojet (a, c, e), and \jets (b, d, f) production 
in the kinematic range specified in table~\ref{tab:kinem}.
 The cross sections are compared to the leading order, $\mathcal{O}(\alpha^3\alpha_s^0)$,
calculation (more details in the caption of figure~\ref{fig:inclxsecsingle}). The \jets sample is additionally compared to a NLO ($\alpha^3\alpha_s$) calculation. 
The bin averaged NLO cross sections are indicated by the squares. }
    \label{fig:xsecsingle1jet}
  \end{center}
  \begin{picture}(0,0)
    \put(22,230){\textsf{\large Photon plus no-Jets}}
    \put(110,230){\textsf{\large Photon plus Jet}}
    \put(63.1,224.7){\textsf{(a)}}
    \put(150.3,224.7){\textsf{(b)}}
    \put(63.1,161.1){\textsf{(c)}}
    \put(150.3,161.1){\textsf{(d)}}
    \put(63.1,99.7){\textsf{(e)}}
    \put(150.3,99.7){\textsf{(f)}}
  \end{picture}
\end{figure}

The measured differential cross sections for the photon plus no-jets and 
photon plus jet selections are 
presented in table~\ref{tab:jets}. They are compared with the LO predictions in figure~\ref{fig:xsecsingle1jet}.
For both samples the LO prediction describes
the shapes of the $d\sigma/dE_T^{\gamma}$ and $d\sigma/d\eta^{\gamma}$ distributions
 reasonably well and  is
 furthest below the data at low $Q^2$, where the $QQ$ term dominates.
All four diagrams in figure~\ref{fig:feynman} contribute to the photon plus no-jets and photon plus jet samples, but the \LL contribution is largely suppressed for the \nojet sample due to the cut on $W_X$.
Since, at leading order ${\cal O}(\alpha_s^0)$, the quark fragmentation contribution to the cross section enters
 only the photon plus no-jets sample,
the observed  excess can not solely be attributed to an underestimation of that contribution.
The cross section for  \jets production is roughly two times higher than 
for  photon plus no-jets.
This is in contrast to the inclusive $ep \rightarrow eX$ cross section, where
topologies with an additional jet are suppressed by $\mathcal{O}(\alpha_s)$.
The similar cross sections for photon events with or without additional jets  can be explained by the fact that 
both topologies correspond to the same order ${\cal O}(\alpha_s^0)$ in perturbative QCD.
\par
In addition, the differential cross sections $d\sigma/dE_T^{\gamma}$, $d\sigma/d\eta^{\gamma}$  and
$d\sigma/dQ^2$
for the photon         plus jet selection
are compared to the NLO prediction (figure~\ref{fig:xsecsingle1jet} right).
On average, the NLO prediction is  higher than 
the LO prediction, in particular at low $Q^2$,
 but is still lower than the data by roughly $35$\%. 
The shapes of all three differential cross sections are described well by the NLO 
prediction.

\section{Conclusions}

The cross section for isolated  photon production $ep \rightarrow e \gamma X$ is measured in deep-inelastic scattering at HERA. Photons with a transverse energy in the range  $3 < E_T^\gamma < 10$~GeV and with pseudorapidity   $-1.2 < \eta^\gamma < 1.8$
are measured in DIS events in the kinematic region  $4<Q^2<150$~GeV$^2$, $y>0.05$ and $W_X>50$~GeV. 
Compared to previous measurements, the range of $Q^2$  is extended from 
$Q^2>35$~GeV$^2$ to $Q^2>4$~GeV$^2$.
The  cross section receives contributions from 
photon radiation by the struck quark ($QQ$), as well as from 
wide angle bremsstrahlung of the initial and final electrons ($LL$) and their interference. 
\par
The data are compared to a LO  ${\cal O}(\alpha^3\alpha_s^0)$ calculation  which is shown to underestimate the measured cross sections by roughly a factor of two. The prediction is  most significantly below the data  at low $Q^2$.
The shapes of the $d\sigma/dE_T^\gamma$ and $d\sigma/d\eta^\gamma$ distributions  are described reasonably well.
The comparison of data to the LO calculation in  bins of $\eta^\gamma$  show that
the difference in normalisation can mainly be attributed to an underestimation of the \QQ contribution.
The data are further compared to predictions from the MC  generators PYTHIA, for the simulation of photons radiated from the quark, and 
RAPGAP for photons radiated from the electron.
 The MC predictions are very similar to the predictions from the LO calculation and also understimate the data, especially at low $Q^2$.
\par
Jet production in events with isolated photons is also studied.
The cross sections for events with no or at least one hadronic jet 
are underestimated by the LO prediction by a similar factor as for the inclusive measurement.
Again the expectations are furthest below the data at low $Q^2$.
The total \jets cross section is roughly double the  \nojet cross section as expected from the calculations. The NLO  ${\cal O}(\alpha^3\alpha_s^1)$ prediction for \jets is  higher than the LO prediction, most significantly at low $Q^2$, but still underestimates the data. The NLO calculation describes the shapes of the differential cross sections   reasonably well.
\par
Further theoretical investigations are needed to understand the
 observed discrepancy between the measurements and the  predictions, including
 for instance the calculation of higher order processes.
\section*{Acknowledgements}

We are grateful to the HERA machine group whose outstanding
efforts have made this experiment possible. 
We thank
the engineers and technicians for their work in constructing and
maintaining the H1 detector, our funding agencies for 
financial support, the
DESY technical staff for continual assistance
and the DESY directorate for support and for the
hospitality which they extend to the non DESY 
members of the collaboration.
We would like to thank Aude Gehrmann-De Ridder, Thomas Gehrmann and Eva Poulsen for providing the LO calculations and many helpful discussions and G. Kramer and H. Spiesberger for the NLO calculation.


\newpage
%
\renewcommand{\arraystretch}{1.15} 
\begin{table}[hhh]
\begin{center}
\begin{tabular}{|l|l|}
\hline
\multicolumn{2}{|c|}{\bf \boldmath{Isolated Photon Cross Section Phase Space } }\\
\hline
\multirow{9}{30mm}{ Inclusive  \\ cross section} & $3 < E_T^\gamma < 10$ GeV \\
   &  $-1.2 < \eta^\gamma < 1.8$ \\
   &  $z=E_T^\gamma/E_T^{photon-jet} > 0.9$ \\
   &  $E_e>10$ GeV \\
   &  $153 < \theta_e < 177^{\circ}$ \\
   &  $4 < Q^2 < 150$ GeV$^2$ \\
   &  $W_X >50$ GeV  \\
   &  $y >0.05$ \\
   &  $\sqrt{s}=319$ GeV \\ \hline
\multirow{4}{30mm}{ Jet definition} & $k_T$ algorithm  with $P_T$-weighted \\
  & recombination scheme~\cite{jetalgo}, $R_0=1$\\
  &  $P_T^{jet} > 2.5 $ GeV\\
  & $-1.0 < \eta^{jet} < 2.1$ (hadronic jet)\\ 
  & $-2.0 < \eta^{photon-jet} < 2.1$ (photon-jet)\\ 
  \hline
\end{tabular}
\caption{ Phase space region in which isolated prompt photon  cross
sections are measured together with the definition of jets. Kinematics are defined in the H1 laboratory frame.}
\label{tab:kinem}
\end{center}
\end{table}
\renewcommand{\arraystretch}{1.35} 
\begin{table}[hhh]
  \begin{center}
    \begin{tabular}{|cc|rrr|c|}
      \hline
      \multicolumn{6}{|c|}{\bf \boldmath H1 Inclusive Isolated Photon Cross Sections}  \\
      \hline
      \hline
      \multicolumn{2}{|c|}{$E_T^\gamma$} & $ d\sigma /d E_T^\gamma$ & stat. & syst. & $f_{had}$ \\
      \multicolumn{2}{|c|}{[GeV]} & \multicolumn{3}{c|}{[pb/GeV]} & \\
      \hline
      $3.0$  &  $4.0$   & $  16.98 $ & $ \pm  1.20 $ & $ ^{+  2.79}_{-  2.61} $ & $0.78$ \\
      $4.0$  &  $6.0$   & $  10.51 $ & $ \pm  0.47 $ & $ ^{+  1.50}_{-  1.86} $ & $0.89$ \\
      $6.0$  &  $10.0$  & $  3.08 $ & $ \pm  0.20 $ & $ ^{+  0.46}_{-  0.60} $ & $0.98$ \\
      \hline
      \hline
      \multicolumn{2}{|c|}{$\eta^\gamma$} & $ d\sigma /d \eta^\gamma$ & stat. & syst. & $f_{had}$ \\
      \multicolumn{2}{|c|}{} & \multicolumn{3}{c|}{[pb]} & \\
      \hline
      $-1.2 $ &$  -0.6 $ & $  26.15 $ & $ \pm  1.67 $ & $ ^{+  3.60}_{-  4.16}   $  & $0.92 $\\
      $-0.6 $ &$  0.2  $&  $  20.69 $ & $ \pm  1.34 $ & $ ^{+  3.53}_{-  3.73}  $  & $0.85 $\\
      $0.2 $ & $ 0.9 $&    $  15.83 $ & $ \pm  0.93 $ & $ ^{+  1.97}_{-  3.25} $  & $0.81 $\\
      $0.9 $&  $1.4 $&     $   9.57 $ & $ \pm  0.87 $ & $ ^{+  1.99}_{-  2.00} $  & $0.80 $\\
      $1.4 $&  $1.8 $&     $   5.50 $ & $ \pm  1.15 $ & $ ^{+  1.04}_{-  1.75} $  & $0.80 $\\
      \hline
      \hline
      \multicolumn{2}{|c|}{$Q^2$} & $ d\sigma /d Q^2$ & stat. & syst. & $f_{had}$ \\
      \multicolumn{2}{|c|}{[GeV$^2$]} & \multicolumn{3}{c|}{[pb/GeV$^2$]} & \\
      \hline
      $4.0 $ &  $10.0$ &$   2.48 $ & $ \pm  0.21 $ & $ ^{+  0.34}_{-  0.41}   $&$ 0.87$ \\
      $10.0$  & $20.0$ &$   1.17 $ & $ \pm  0.07 $ & $ ^{+  0.19}_{-  0.21}   $&$ 0.83$ \\
      $20.0$  & $40.0$ &$   0.52 $ & $ \pm  0.03 $ & $ ^{+  0.07}_{-  0.10}   $&$ 0.81 $\\
      $40.0$ &  $80.0$ &$    0.235 $ & $ \pm  0.013 $ & $ ^{+  0.033}_{-  0.048}  $ &$ 0.83$ \\
      $80.0$ &  $150.$ &$     0.063 $ & $ \pm  0.006 $ & $ ^{+  0.009}_{-  0.012} $ &$ 0.87$ \\
      \hline
    \end{tabular}
    \caption{Differential cross sections for inclusive isolated photon production
in the kinematic range specified in table~\ref{tab:kinem}.
$f_{had}$ denotes the hadronisation correction factor applied to the LO calculation.}
    \label{tab:inclusive}
  \end{center}
\end{table}

\begin{table}[hhh]
  \begin{center}
    \begin{tabular}{|cc|rrr|c|}      
      \hline
      \multicolumn{6}{|c|}{\bf \boldmath H1 Inclusive Isolated Photon Cross Sections} \\
      \hline
      \hline
      \multicolumn{2}{|c|}{$E_T^\gamma$} & $ d\sigma /d E_T^\gamma$ & stat. & syst. & $f_{had}$ \\
      \multicolumn{2}{|c|}{[GeV]} & \multicolumn{3}{c|}{[pb/GeV]} & \\
      \hline
      \hline
      \multicolumn{6}{|c|}{$-1.2< \eta^\gamma < -0.6 $} \\
      \hline
      $3.0 $ & $ 4.0  $ &  $  4.86 $ & $ \pm  0.67 $ & $ ^{+  0.88}_{-  0.63}   $& $0.86$\\
      $4.0 $ & $ 6.0  $ &  $   3.46 $ & $ \pm  0.28 $ & $ ^{+  0.48}_{-  0.66}  $& $0.96$ \\
      $6.0 $ & $ 10.0 $ &  $    0.98 $ & $ \pm  0.12 $ & $ ^{+  0.13}_{-  0.23}  $ &$ 1.00$ \\
      \hline
      \hline
      \multicolumn{6}{|c|}{$-0.6< \eta^\gamma < 0.2 $} \\
      \hline
      $3.0 $ & $ 4.0  $ & $    5.81 $ & $ \pm  0.75 $ & $ ^{+  1.27}_{-  1.20}  $    & $0.76$ \\
      $4.0 $ & $ 6.0  $ & $    3.20 $ & $ \pm  0.28 $ & $ ^{+  0.56}_{-  0.65}  $  & $0.88 $\\
      $6.0 $ & $ 10.0 $ & $    1.09 $ & $ \pm  0.13 $ & $ ^{+  0.15}_{-  0.17}  $   &$ 0.99$ \\
      \hline
      \hline
      \multicolumn{6}{|c|}{$0.2< \eta^\gamma < 0.9 $} \\
      \hline
      $3.0 $ & $ 4.0   $& $   3.94 $ & $ \pm  0.51 $ & $ ^{+  0.59}_{-  0.72}  $ & $0.72$ \\
      $4.0 $ & $ 6.0   $& $    2.39 $ & $ \pm  0.16 $ & $ ^{+  0.28}_{-  0.51} $ & $0.84$ \\
      $6.0 $ & $ 10.0  $& $    0.59 $ & $ \pm  0.06 $ & $ ^{+  0.09}_{-  0.14} $ & $0.96$ \\
      \hline
      \hline
      \multicolumn{6}{|c|}{$0.9< \eta^\gamma < 1.4 $} \\
      \hline
      $3.0 $ & $ 4.0 $  &  $   1.66 $ & $ \pm  0.31 $ & $ ^{+  0.22}_{-  0.31}  $   & $0.69$ \\
      $4.0 $ & $ 6.0 $  &  $   0.82 $ & $ \pm  0.12 $ & $ ^{+  0.21}_{-  0.16} $   & $0.82$ \\
      $6.0 $ & $ 10.0$  &  $    0.37 $ & $ \pm  0.05 $ & $ ^{+  0.10}_{-  0.09} $   & $0.96 $\\
      \hline
      \hline
      \multicolumn{6}{|c|}{$1.4< \eta^\gamma < 1.8 $} \\
      \hline
      $3.0 $ & $ 4.0  $ & $ 0.72 $ & $ \pm  0.28 $ & $ ^{+  0.23}_{-  0.21}  $  & $0.70$ \\
      $4.0 $ & $ 6.0  $ & $  0.64 $ & $ \pm  0.16 $ & $ ^{+  0.09}_{-  0.22} $  & $0.81$ \\
      $6.0 $ & $ 10.0 $ & $  0.049 $ & $ \pm  0.045 $ & $ ^{+  0.009}_{-  0.016} $  & $0.94$ \\
      \hline
    \end{tabular}
    \caption{
Differential cross sections for inclusive isolated photon production
    $d\sigma/dE_{T}^{\gamma}$  in different
    $\eta^{\gamma}$ bins  corresponding to the wheel structure of the LAr
    calorimeter (see text).
The kinematic range is defined in table~\ref{tab:kinem}.
 $f_{had}$ denotes the hadronisation correction factor applied to the LO calculation.}
    \label{tab:double}
  \end{center}
\end{table}

\begin{table}[hhh]
  \begin{center}
    \begin{tabular}{|cc|rrr|c|}
      \hline
      \multicolumn{6}{|c|}{\bf \boldmath H1 Inclusive Isolated Photon Cross Sections}  \\
      \multicolumn{6}{|c|}{\bf \boldmath for $Q^2>40$ GeV$^2$}  \\
      \hline
      \hline
      \multicolumn{2}{|c|}{$E_T^\gamma$} & $ d\sigma /d E_T^\gamma$ & stat. & syst. & $f_{had}$ \\
      \multicolumn{2}{|c|}{[GeV]} & \multicolumn{3}{c|}{[pb/GeV]} & \\
      \hline
      $3.0  $ & $  4.0   $ & $   3.70 $ & $ \pm  0.39 $ & $ ^{+  0.63}_{-  0.59}  $ & $ 0.80 $\\
      $4.0  $ & $  6.0   $ & $   2.53 $ & $ \pm  0.23 $ & $ ^{+  0.43}_{-  0.43}  $ & $ 0.87 $\\
      $6.0  $ & $  10.0  $ & $    1.30 $ & $ \pm  0.15 $ & $ ^{+  0.19}_{-  0.27} $ & $ 0.96 $\\
      \hline
      \hline
      \multicolumn{2}{|c|}{$\eta^\gamma$} & $ d\sigma /d \eta^\gamma$ & stat. & syst. & $f_{had}$ \\
      \multicolumn{2}{|c|}{} & \multicolumn{3}{c|}{[pb]} & \\
      \hline
      $-1.2  $ & $  -0.6  $ & $   9.61 $ & $ \pm  1.00 $ & $ ^{+  1.48}_{-  2.02}  $ & $ 0.97 $\\
      $-0.6  $ & $  0.2  $ & $   5.13 $ & $ \pm  0.59 $ & $ ^{+  0.92}_{-  0.97}  $ & $ 0.86 $\\
      $0.2  $ & $  0.9 $ & $   3.49 $ & $ \pm  0.32 $ & $ ^{+  0.45}_{-  0.72}   $ & $ 0.78 $\\
      $0.9 $ & $  1.4 $ & $    2.37 $ & $ \pm  0.33 $ & $ ^{+  0.49}_{-  0.47}  $ & $ 0.76 $\\
      $1.4 $ & $  1.8 $ & $    1.12 $ & $ \pm  0.51 $ & $ ^{+  0.21}_{-  0.34}  $ & $ 0.73 $\\
      \hline
    \end{tabular}
    \caption{Differential cross sections for inclusive isolated photon production
in the kinematic range specified in table~\ref{tab:kinem} and $40<Q^{2}<150$ GeV$^2$.
 $f_{had}$ denotes the hadronisation correction factor applied to the LO calculation.}
    \label{tab:highQ2}
  \end{center}
\end{table}

\begin{table}[hhh]
  \begin{center}
    \begin{tabular}{|cc|rrr|c||rrr|c|}
      \hline
      \multicolumn{2}{|c|}{} & \multicolumn{4}{c|}{\bf \boldmath H1 Photon plus no-Jets} &\multicolumn{4}{c|}{\bf \boldmath H1 Photon plus Jet} \\
      \hline
      \multicolumn{2}{|c|}{$E_T^\gamma$} & $d\sigma /d E_T^\gamma$ & stat. & syst. & $f_{had}$& $ d\sigma /d E_T^\gamma$ & stat. & syst. & $f_{had}$ \\
      \multicolumn{2}{|c|}{[GeV]} & \multicolumn{3}{c|}{[pb/GeV]} & & \multicolumn{3}{c|}{[pb/GeV]} & \\
      \hline
      $3.0  $ & $  4.0   $ & $  
    8.10 $ & $ \pm  0.93 $ & $ ^{+  1.82}_{-  1.48}$ & $ 0.75 $ & $
    8.85 $ & $ \pm  0.70 $ & $ ^{+  1.37}_{-  1.38}$ & $ 0.82$\\
      $4.0  $ & $  6.0   $ & $
    3.79 $ & $ \pm  0.29 $ & $ ^{+  0.69}_{-  0.77}$ & $ 0.91$ & $
    6.65 $ & $ \pm  0.35 $ & $ ^{+  0.92}_{-  1.15}$ & $ 0.89 $\\
      $6.0  $ & $  10.0  $ & $
    0.77 $ & $ \pm  0.10 $ & $ ^{+  0.14}_{-  0.18}$ & $1.10 $& $
    2.35 $ & $ \pm  0.17 $ & $ ^{+  0.35}_{-  0.46} $ & $ 0.97 $\\
      \hline
      \hline
      \multicolumn{2}{|c|}{$\eta^\gamma$} & $ d\sigma /d \eta^\gamma$ & stat. & syst. & $f_{had}$& $ d\sigma /d \eta^\gamma$ & stat. & syst. & $f_{had}$ \\
      \multicolumn{2}{|c|}{} & \multicolumn{3}{c|}{[pb]} & & \multicolumn{3}{c|}{[pb]} & \\
      \hline
      $-1.2  $ & $  -0.6  $ & $ 
   9.30 $ & $ \pm  1.07 $ & $ ^{+  1.53}_{-  1.72}$ & $ 0.88 $ & $ 
  16.61 $ & $ \pm  1.20 $ & $ ^{+  2.67}_{-  2.64}$ & $ 0.97 $\\
      $-0.6  $ & $  0.2  $ & $   
  8.46 $ & $ \pm  0.95 $ & $ ^{+  1.73}_{-  1.84}$ & $ 0.81 $ & $
   12.32 $ & $ \pm  0.90 $ & $ ^{+  2.15}_{-  2.19}   $ & $ 0.88 $\\
      $0.2  $ & $  0.9 $ & $ 
 5.98 $ & $ \pm  0.71 $ & $ ^{+  1.16}_{-  1.34}$ & $ 0.82 $ & $
 9.94 $ & $ \pm  0.59 $ & $ ^{+  1.16}_{-  2.03}   $ & $ 0.81 $\\
      $0.9 $ & $  1.4 $ & $   
  2.57 $ & $ \pm  0.47 $ & $ ^{+  0.64}_{-  0.61}$ & $ 0.85 $ & $ 
  6.99 $ & $ \pm  0.73 $ & $ ^{+  1.38}_{-  1.40}$ & $ 0.79 $\\
      $1.4 $ & $  1.8 $ & $ 
   2.40 $ & $ \pm  0.73 $ & $ ^{+  0.78}_{-  0.74}$ & $ 0.91 $ & $ 
   3.22 $ & $ \pm  0.85 $ & $ ^{+  0.61}_{-  1.01}$ & $ 0.77 $\\
      \hline
      \hline
      \multicolumn{2}{|c|}{$Q^2$} & $d\sigma /d Q^2$ & stat. & syst. & $f_{had}$ & $ d\sigma /d Q^2$ & stat. & syst. & $f_{had}$ \\
      \multicolumn{2}{|c|}{[GeV$^2$]} & \multicolumn{3}{c|}{[pb/GeV$^2$]} & & \multicolumn{3}{c|}{[pb/GeV$^2$]} & \\
      \hline
      $4.0  $ & $  10.0  $ & $
   1.09 $ & $ \pm  0.16 $ & $ ^{+  0.21}_{-  0.22}$ & $ 0.88 $ & $
  1.39 $ & $ \pm  0.13 $ & $ ^{+  0.20}_{-  0.22}$ & $ 0.87 $\\
      $10.0  $ & $ 20.0  $ & $ 
   0.44 $ & $ \pm  0.05 $ & $ ^{+  0.09}_{-  0.09}$ & $ 0.81 $ & $
   0.76 $ & $ \pm  0.05 $ & $ ^{+  0.11}_{-  0.14}$ & $ 0.86 $\\
      $20.0  $ & $ 40.0 $ & $ 
   0.21 $ & $ \pm  0.02 $ & $ ^{+  0.04}_{-  0.05} $ & $ 0.80 $ & $
 0.31 $ & $ \pm  0.02 $ & $ ^{+  0.04}_{-  0.06}$ & $ 0.83 $\\
      $40.0 $ & $  80.0 $ & $ 
   0.071 $ & $ \pm  0.008 $ & $ ^{+  0.012}_{-  0.017}$ & $ 0.81 $ & $
  0.162 $ & $ \pm  0.010 $ & $ ^{+  0.024}_{-  0.033}$ & $ 0.84 $\\
      $80.0 $ & $  150. $ & $ 
   0.021 $ & $ \pm  0.004 $ & $ ^{+  0.005}_{-  0.007}$ & $ 0.88 $ & $
   0.040 $ & $ \pm  0.005 $ & $ ^{+  0.005}_{-  0.007}$ & 0.89\\
      \hline
    \end{tabular}
    \caption{Differential cross sections for  the production
of isolated photons accompanied by no  or at least one  hadronic jet
in the kinematic range specified in table~\ref{tab:kinem}.
$f_{had}$ denotes the hadronisation correction factor applied to the LO and the NLO calculation.}
    \label{tab:jets}
  \end{center}
\end{table}
 \clearpage


\end{document}